\documentclass[nofootinbib,amsmath,twocolumn,notitlepage,preprintnumbers]{revtex4-1}
\usepackage{multirow}
\usepackage{amssymb, esvect, amsmath, graphicx, latexsym, amsthm, slashed, eso-pic}
\usepackage{xcolor}
\usepackage{hyperref}
\usepackage{amsmath}

\newcommand{\beq}{\begin{equation}} \newcommand{\eeq}{\end{equation}}
\newcommand{\bea}{\begin{eqnarray}} \newcommand{\eea}{\end{eqnarray}}

\def\lsim{\mathrel{\raise.3ex\hbox{$<$\kern-.75em\lower1ex\hbox{$\sim$}}}}
\def\gsim{\mathrel{\raise.3ex\hbox{$>$\kern-.75em\lower1ex\hbox{$\sim$}}}}

\newcommand{\be}{\begin{eqnarray}}
\newcommand{\ee}{\end{eqnarray}}

\newcommand{\benum}{\begin{enumerate}}
\newcommand{\eenum}{\end{enumerate}}
\newcommand{\bi}{\begin{itemize}}
\newcommand{\ei}{\end{itemize}}

\begin{document}

\preprint{FERMILAB-PUB-21-088-T}

\title{The 511 keV Excess and Primordial Black Holes}

\author{Celeste Keith$^{b,c}$}
\thanks{ORCID: https://orcid.org/0000-0002-3004-0930}

\author{Dan Hooper$^{a,b,c}$}
\thanks{ORCID: http://orcid.org/0000-0001-8837-4127}

\affiliation{$^a$Fermi National Accelerator Laboratory, Theoretical Astrophysics Group, Batavia, IL, USA}
\affiliation{$^b$University of Chicago, Kavli Institute for Cosmological Physics, Chicago IL, USA}
\affiliation{$^c$University of Chicago, Department of Astronomy and Astrophysics, Chicago IL, USA}

\date{\today}

\begin{abstract}
An excess of 511 keV photons has been detected from the central region of the Milky Way. It has been suggested that the positrons responsible for this signal might be produced through the Hawking evaporation of primordial black holes. After evaluating the constraints from INTEGRAL, COMPTEL, and Voyager 1, we find that black holes in mass range of $\sim$\,$(1-4)\times10^{16}$ g could potentially produce this signal if they make up a small fraction of the total dark matter density. Proposed MeV-scale gamma-ray telescopes such as AMEGO or e-ASTROGAM should be able to test this class of scenarios by measuring the diffuse gamma ray emission from the Milky Way's inner halo.


\end{abstract}

\maketitle

The INTEGRAL satellite has detected an excess of 511 keV photons from the inner Milky Way relative to astrophysical expectations. This signal consists of a flux of (1.07 $\pm$ 0.03) $\times$10$^{-3}$ photons cm$^{-2}$ s$^{-1}$, requiring the injection of $\sim$\,$2 \times 10^{43}$ positrons per second~\cite{Weidenspointner:2004my, Churazov:2004as, Weidenspointner:2007rs, Jean:2005af, Weidenspointner:2008zz, Kierans:2019aqz, Prantzos:2005pz}. A variety of potential astrophysical sources for this signal have been proposed, including type Ia supernovae~\cite{Kalemci:2006bz}, gamma-ray bursts~\cite{Casse:2003fh,Bertone:2004ek}, microquasars~\cite{Guessoum:2006fs}, low-mass X-ray binaries~\cite{Bartels:2018eyb}, and neutron star mergers~\cite{Takhistov:2019zyb,Fuller:2018ttb}. However, given the challenges involved in explaining the observed characteristics of this signal (for a review, see Ref.~\cite{Prantzos:2010wi}), a number of more exotic scenarios have also been put forth. In particular, the annihilations of MeV-scale dark matter particles have been considered in detail within this context~\cite{Boehm:2003bt,Huh:2007zw,Hooper:2008im,Khalil:2008kp}, although this possibility is constrained by measurements of the damping tail of the cosmic microwave background~\cite{Wilkinson:2016gsy} (see, however, Refs.~\cite{Ema:2020fit, Escudero:2018mvt, Sabti:2019mhn}). 
Explanations featuring decaying~\cite{Hooper:2004qf,Cembranos:2008bw,Craig:2009zv}, or upscattering~\cite{Finkbeiner:2007kk,Pospelov:2007xh,Cline:2010kv,Cline:2012yx} dark matter particles also remain potentially viable. Other exotic scenarios involving Q-balls~\cite{Kasuya:2005ay}, pico-charged particles~\cite{Farzan:2017hol,Farzan:2020llg}, quark nuggets~\cite{Lawson:2016mpu}, or 
unstable MeV-scale states produced in supernovae~\cite{Davoudiasl:2009ud} have also been discussed within this context.

Another possibility that we will consider here is that the excess positrons could be produced through the Hawking evaporation~\cite{Hawking:1974sw,Gibbons:1977mu} of a population of primordial black holes concentrated in the Inner Galaxy~\cite{Frampton:2005fk,Bambi:2008kx,Cai:2020fnq} (see also, Refs.~\cite{Laha:2019ssq,DeRocco:2019fjq}). As we will show, black holes with masses in the range of $m_{\rm BH} \sim (1-4) \times 10^{16}$ g could produce the required flux of positrons while remaining consistent with all existing constraints, including those from the COMPTEL, INTEGRAL, and Voyager 1 satellites~\cite{Laha:2020ivk,Coogan:2020tuf,Boudaud:2018hqb}. For other constraints, see Refs.~\cite{Carr:2009jm, Carr:2020gox, Dasgupta:2019cae}.

In this letter, we explore the possibility that primordial black holes could be the source of the positrons responsible for INTEGRAL's 511 keV excess. After identifying the regions of parameter space that can accommodate this signal, we calculate the gamma-ray spectrum from these black holes (including direct Hawking radiation, photons from the inflight annihilation of positrons, and final-state radiation), and derive constraints based on data from the gamma-ray telescopes INTEGRAL and COMPTEL. In scenarios in which black holes can produce the observed 511 keV signal, we find that the number density of black holes in the local halo is $\sim$\,$10^{12} \, {\rm pc}^{-3}$, corresponding to a median instantaneous distance of $\sim$\,10 AU to the nearest such black hole, well within the boundaries of our Solar System. Given the significant velocities of any primordial black holes in the Milky Way's halo, we should expect individual black holes to regularly pass through the inner Solar System. Even at such close distances, however, it would be very challenging to detect the Hawking radiation from an individual black hole in this mass range. On the other hand, it should be possible to definitively test this class of scenarios by characterizing the diffuse MeV-scale gamma-ray emission from the halo of the Milky Way with proposed gamma-ray telescopes such as AMEGO or \mbox{e-ASTROGAM}.

\begin{figure*}[htp]
\centering
\includegraphics[width=.5\textwidth, height = .25\textwidth]{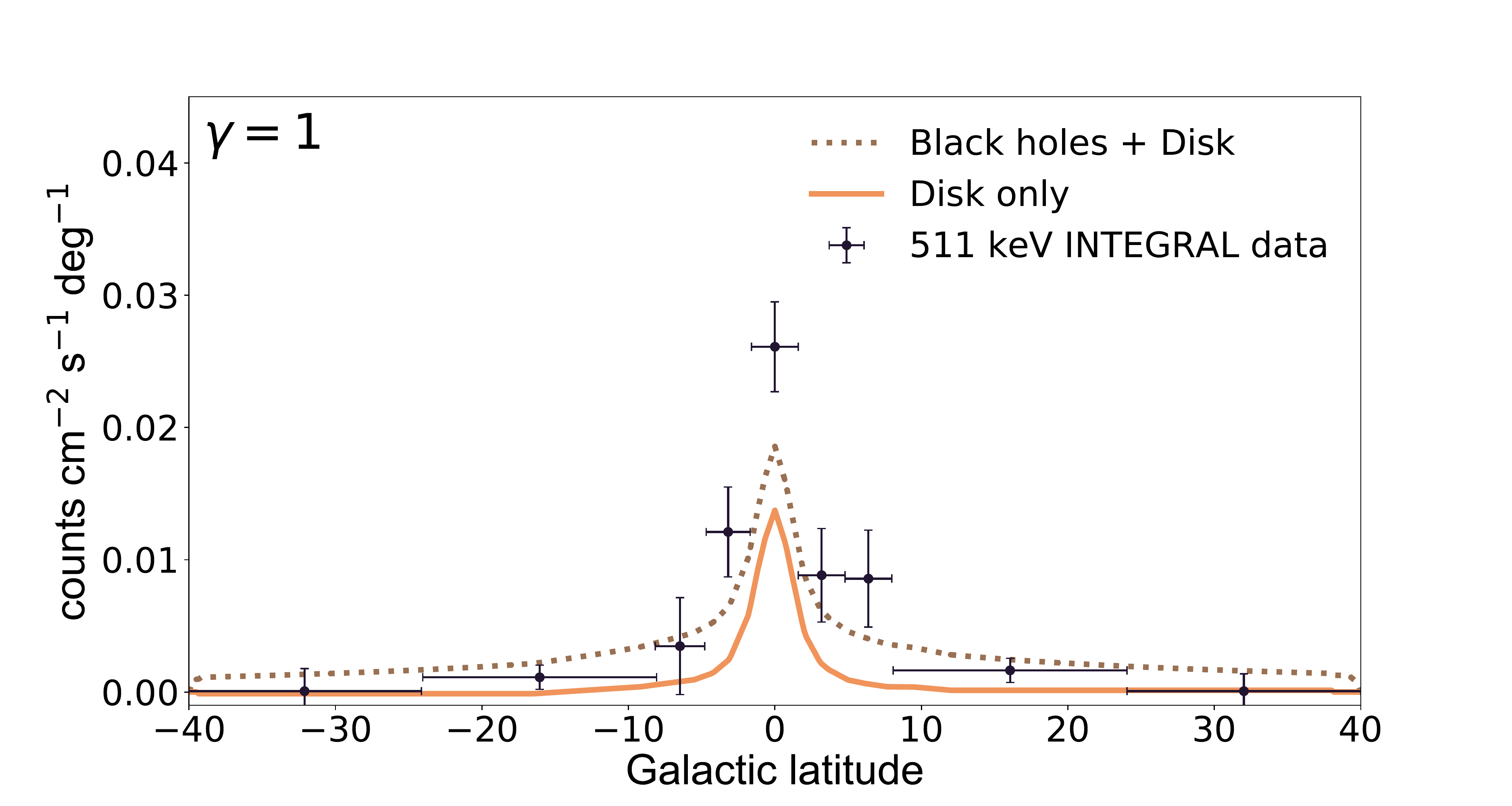}\hfill
\includegraphics[width=.5\textwidth, height = .25\textwidth]{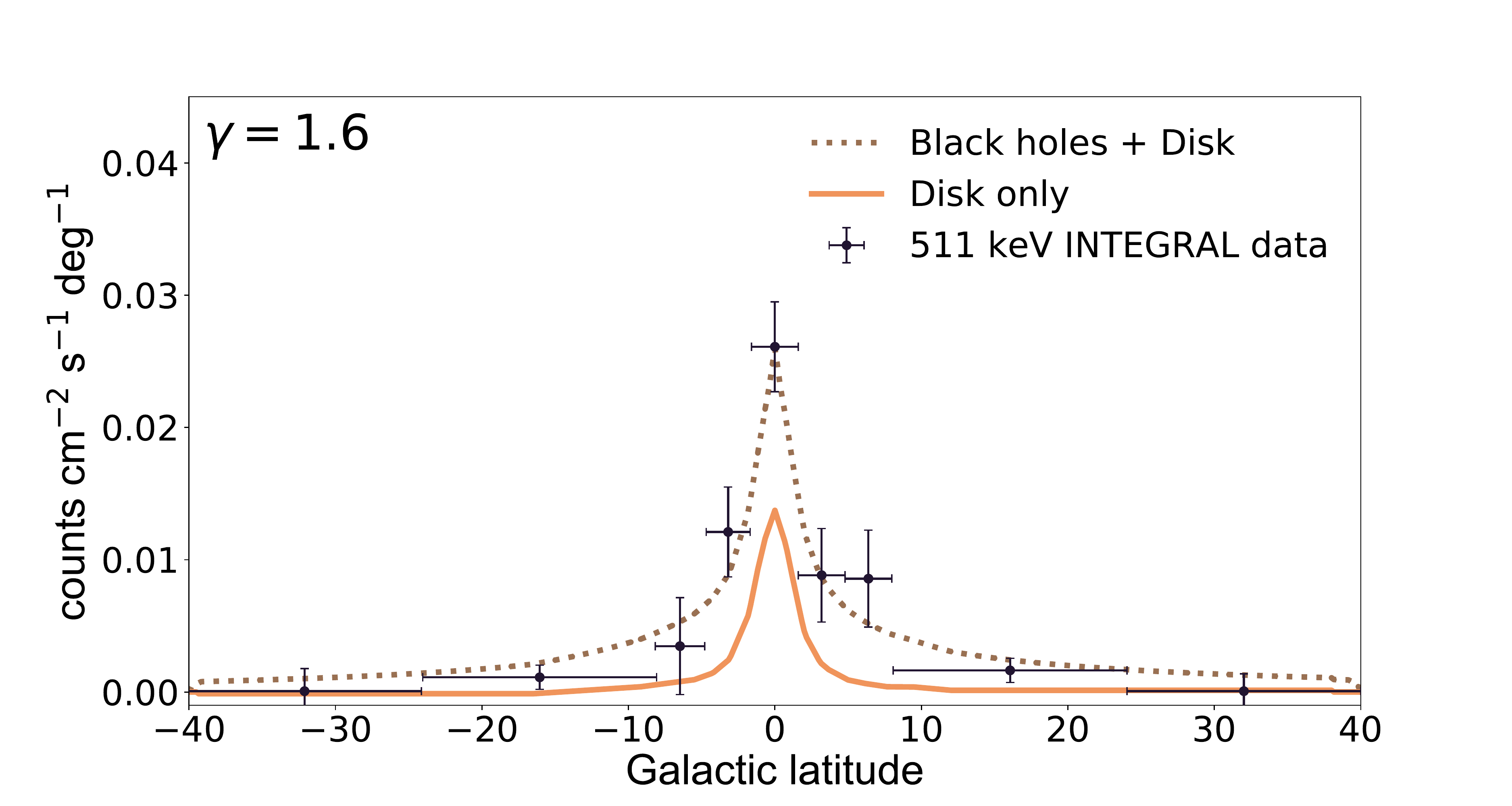}\hfill \\
\includegraphics[width=.5\textwidth, height = .25\textwidth]{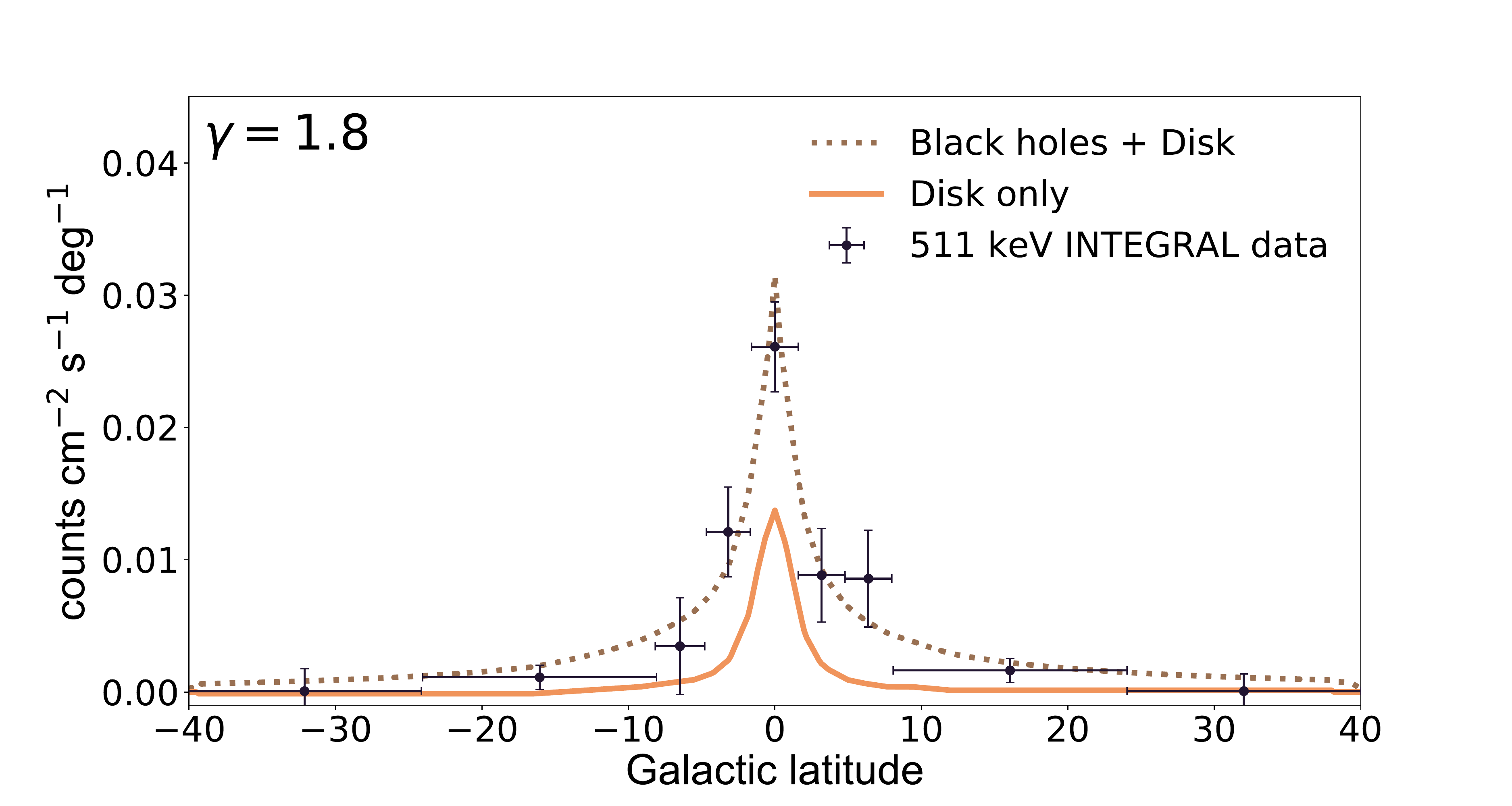}\hfill
\includegraphics[width=.5\textwidth, height = .25\textwidth]{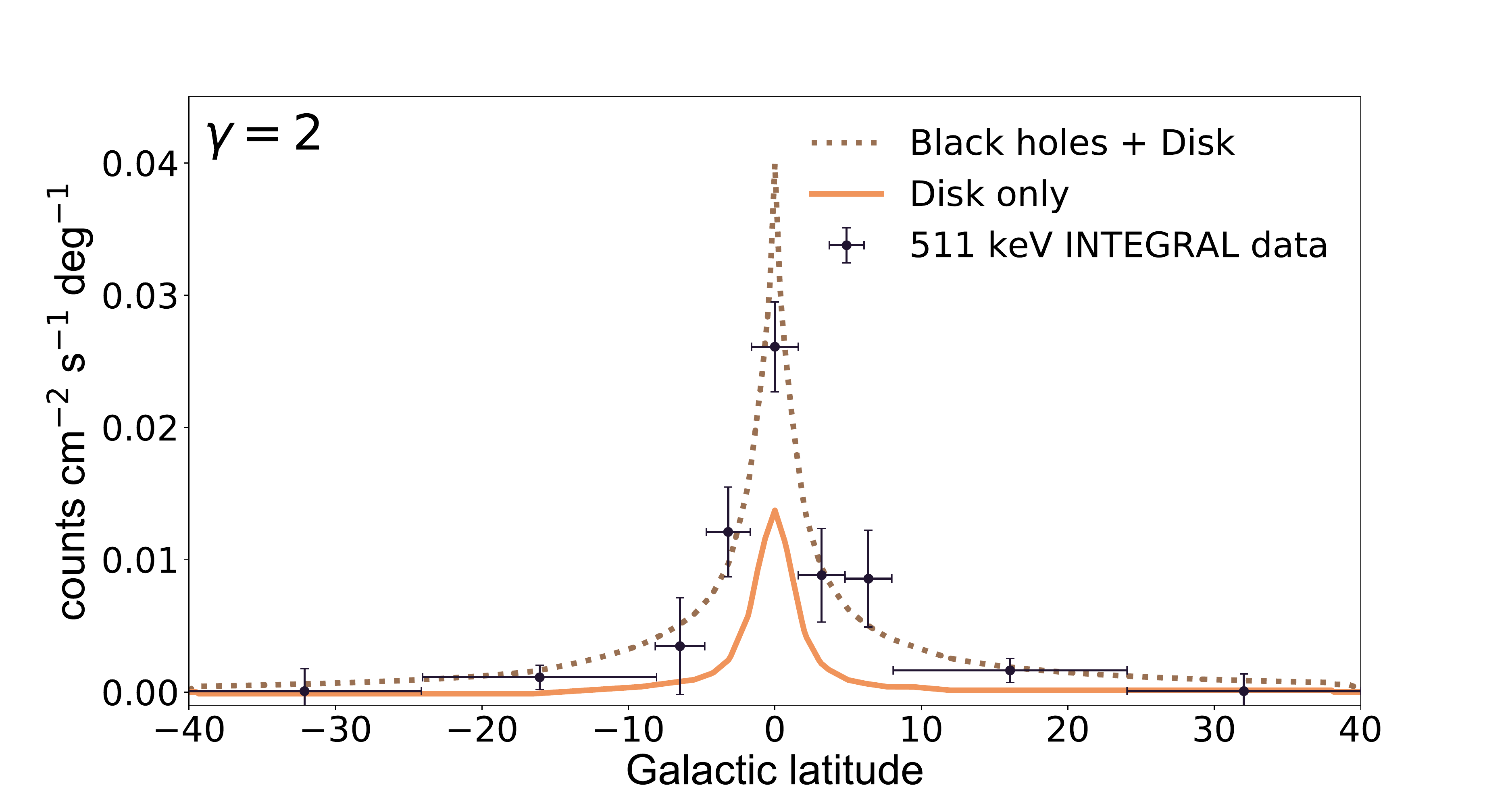}
\caption{The predicted flux and angular profile of 511 keV photons, averaged over $-8^{\circ}< l <+8^{\circ}$, and compared to the measurements of the INTEGRAL satellite~\cite{Bouchet:2010dj}. Results are shown for four choices of the density profile's inner slope, $\gamma$. In each frame, we have selected values of $m_{\rm BH}$ and $f_{\rm DM}$ which provide the best possible normalization to this data. The solid curves represent an estimate for the contribution from astrophysical sources in the Galactic Disk~\cite{Robin:2004qd}, while the dashed curves correspond to the total contribution from disk sources and primordial black holes.}
\label{many_gammas_511}
\end{figure*}

Black holes radiate all particle species lighter than or comparable to their temperature, which is related to the mass of the black hole:
\be \label{masstemp}
T_{\rm BH} = \frac{M_{\text{Pl}}^2}{8 \pi m_{\text{BH}}} \approx 1.05 \, {\rm MeV}  \, \bigg(\frac{10^{16} \, {\rm g}}{m_{\rm BH}}\bigg).
\ee
This radiation causes a black hole\footnote{For context, the Schwarzschild radius of a black hole is given by $r_s=2m_{\rm BH}/M^{2}_{\rm Pl} \simeq 1.5 \times 10^{-12} \, {\rm cm} \times (m_{\rm BH}/10^{16} {\rm g})$} to lose mass at the following rate:
\be \label{dmdt}
\!\begin{aligned}
 \frac{dm_{\textrm{BH}}}{dt} &= -\frac{\mathcal{G}g_{*,H}(m_{\textrm{BH}})M_{\textrm{Pl}}^2}{30720\pi m_{\textrm{BH}}^2} \\
& \approx -8.2 \times 10^{-7} \,\textrm{g/s}  \left(\frac{g_{*,H}}{10.92} \right) \left(\frac{10^{16}\textrm{g}}{m_{\textrm{BH}}} \right)^2, \\
\end{aligned} 
\ee
where $\mathcal{G} \approx$ 3.8 is the appropriate greybody factor, $M_{\text{Pl}}$ is the Planck mass, and $g_{*,H}$ counts the number of (spin-weighted) degrees-of-freedom with $m \ll T_{\rm BH}$, receiving a contribution of 6 from neutrinos, 4 from electrons, 0.82 from photons, and 0.1 from gravitons~\cite{MacGibbon:1990zk,MacGibbon:1991tj}. Integrating this expression, one finds that a black hole with an initial mass of $m_{\rm BH} \sim 5 \times 10^{14}$g will evaporate in a time equal to the age of the universe. 

The spectrum of Hawking radiation from an individual black hole is given by~\cite{Page:1976df}:
\be \label{page}
\frac{dN}{dE}(m_{\text{BH}}, E) = \frac{1}{2\pi^{2}}\frac{E^{2}\sigma(m_{\text{BH}},E)}{e^{E/T}\pm1},
\ee
where both the sign in the denominator and the absorption cross section, $\sigma$, depend on the spin of the particle species being radiated. For fermions (bosons), the sign is positive (negative). In the high-energy limit ($E \gg T$), the absorption cross section approaches $\sigma \simeq 27 \pi m^2_{\rm BH}/M^4_{\rm Pl}$ for all particle species. In the opposite limit, this cross section depends both on the particle species in question, and on the energy of those particles. In performing our calculations, we do not rely on any limiting behavior, but instead use the full spectra as presented in Ref.~\cite{Page:1976df}.

\begin{figure*}
\includegraphics[width=3.4in,angle=0]{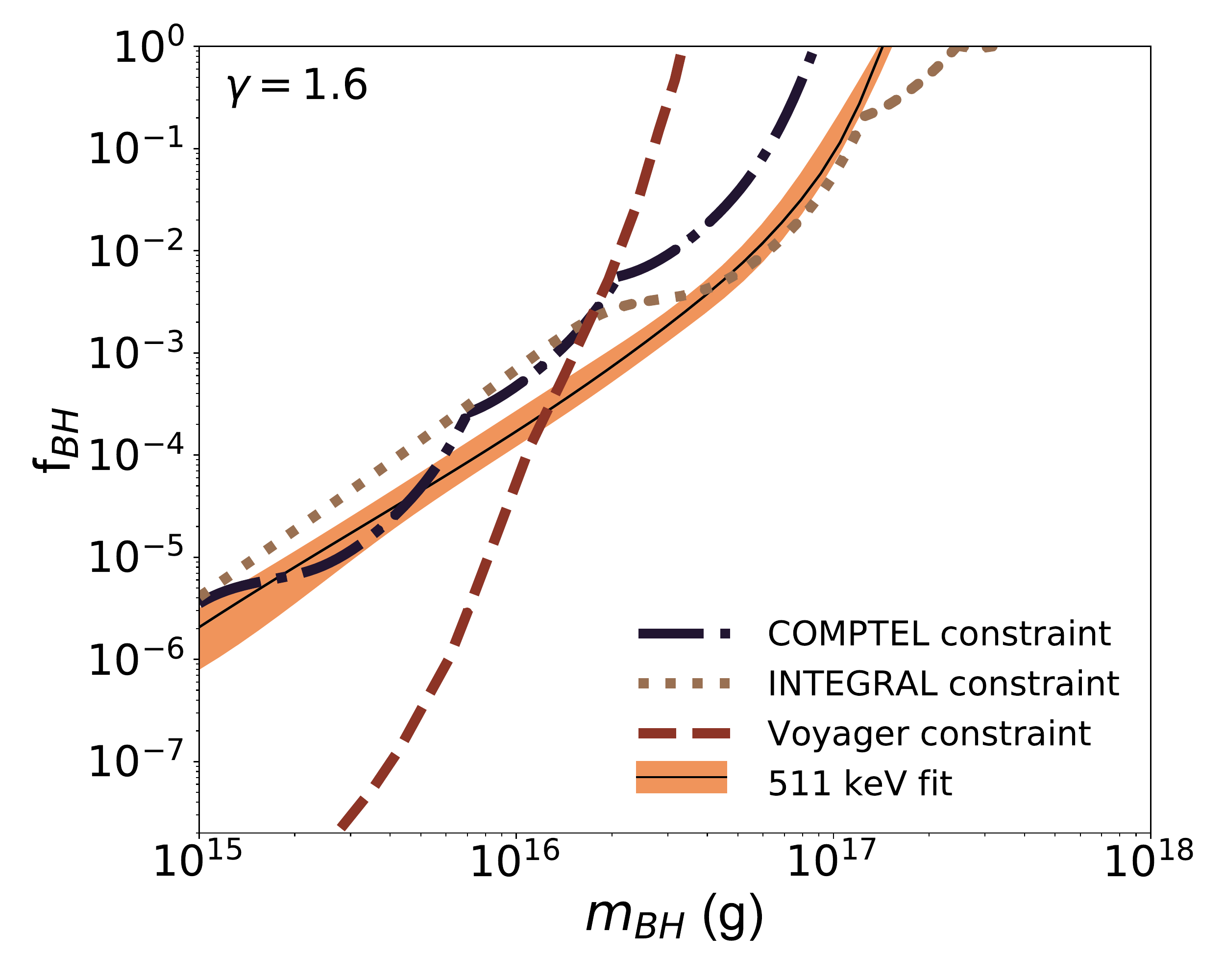}
\includegraphics[width=3.4in,angle=0]{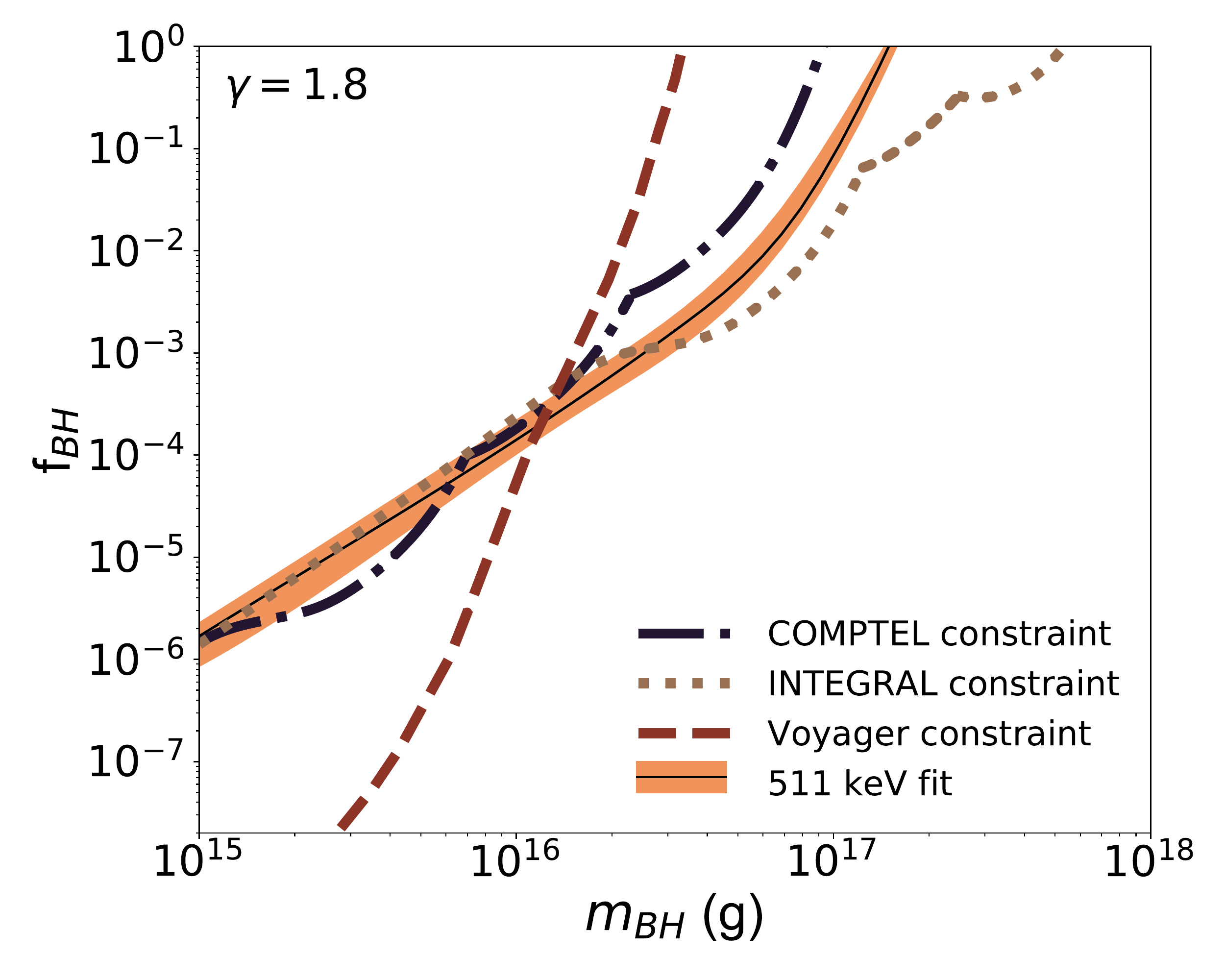}
\caption{In each frame, the orange band denotes the 2$\sigma$ region of parameter space in which the 511 keV excess observed by INTEGRAL could be produced through the Hawking evaporation of primordial black holes. Also shown are the constraints on this parameter space derived from the measurements of the INTEGRAL, COMPTEL, and Voyager 1 satellites. Black holes with masses of $\sim (1-4) \times 10^{16}$ g could produce the observed excess if they constitute a small fraction of the total dark matter and are distributed according to a halo profile with a very steep inner slope, $\gamma \simeq 1.6-1.8$.}
\label{18gamma}
\end{figure*}

In order to identify the parameter space in which primordial black holes could produce the observed 511 keV excess, we calculate the flux and spatial distribution of the positrons injected from these objects and compare this to the intensity and morphology of the 511 keV signal as reported in Ref.~\cite{Bouchet:2010dj}. The flux of 511 keV photons from black holes near the Galactic Center observed over a region of solid angle, $\Delta \Omega$, is given by:
\begin{align} \label{jfactor}
F_{511}(\Delta \Omega) &= \frac{L_{511}(m_{\rm BH})}{4\pi} \int_{\Delta \Omega} \int_{los} n_{\textrm{BH}}(l, \Omega) \,dl \, d\Omega, \\
& \approx \frac{0.55 \,L_{e^+}(m_{\rm BH})\, f_{\rm BH}}{4\pi m_{\textrm{BH}}} \int_{\Delta \Omega} \int_{los} \rho_{\textrm{DM}}(l, \Omega) \,dl \, d\Omega, \nonumber
\end{align}
where $n_{\rm BH}$ is the number density of black holes, $L_{511}$ is the number of 511 keV photons produced by a given black hole per unit time, and the integrals are performed over solid angle and the line-of-sight, $l$. In the second line, we have written the number density of black holes in terms of the dark matter density, $n_{\rm BH} = f_{\rm BH} \,\rho_{\rm DM} /m_{\rm BH}$, where $f_{\rm BH}$ is the fraction of the total dark matter that consists of black holes. When a black hole emits a positron, it can either directly annihilate to produce two 511 keV photons, or form a positronium bound state with an electron, which results 25\% of the time in two 511 keV photons, and 75\% of the time in three photons, each with $E_{\gamma} <$ 511 keV. Given that observations indicate that the positronium fraction in the interstellar medium of the Milky Way is $f = 0.967 \pm 0.022$ \cite{Jean:2005af}, the number of 511 keV photons produced per positron is $2(1-f)+2f/4 \approx 0.55$. 

For the dark matter distribution, we adopt a generalized Navarro-Frenk-White (NFW) halo profile \cite{Navarro:1995iw,Navarro:1996gj}:
\begin{align}
\rho_{\rm DM} =\frac{\rho_0}{(r/R_s)^{\gamma} \, [1+(r/R_s)]^{3-\gamma}},
\end{align}
where $r$ is the distance from the Galactic Center. Throughout this study, we have adopted a scale radius of $R_s=20$ kpc and have normalized $\rho_0$ such that the local density (at $r=8.25$ kpc) is 0.4 GeV/cm$^3$. We take the inner slope of this profile, $\gamma$, to be a free parameter.

In Fig.~\ref{many_gammas_511}, we have plotted the 511 keV emission predicted from primordial black holes as a function of galactic latitude (averaging over $-8^{\circ}< l <+8^{\circ}$), for four choices of the density profile's inner slope, $\gamma$. In each frame, we compare the predicted profile of this emission to the measurements by INTEGRAL~\cite{Bouchet:2010dj}. In addition to the 511 keV emission resulting from positrons radiated from black holes, we have included an estimate for the contribution from astrophysical positron emission in the disk, as described in Ref.~\cite{Robin:2004qd}. Given the uncertainties regarding the magnitude of the astrophysical contribution, we have allowed its normalization to float from its default value by up to a factor of two. From this figure, it is clear that a very steep density profile is required if primordial black holes are to produce the observed intensity of the 511 keV emission from the inner few degrees around the Galactic Center (see also, Refs.~\cite{Vincent:2012an,Ascasibar:2005rw}). Formally, our fit to this data favors $\gamma = 2.2 \pm 0.6$ (at $2\sigma$), although a somewhat larger range could plausibly be accommodated if all of the related uncertainties were taken into account. While this range is well above that typically favored by numerical simulations of cold dark matter ($\gamma \sim 1-1.4$~\cite{gnedin2011halo, Gnedin:2004cx, Governato_2012, Kuhlen:2012qw, Weinberg:2001gm, Weinberg:2006ps, Sellwood:2002vb, Valenzuela:2002np, Colin:2005rr, Scannapieco_2012, Calore:2015oya, Schaller:2014uwa, DiCintio:2014xia, DiCintio:2013qxa, Schaller:2015mua, Bernal:2016guq}), the lower portion of the range favored by our fit could be potentially viable if adiabatic contraction is efficient (see, for example, Ref.~\cite{Gnedin:2011uj}). In light of this, we chose to focus on the lowest portion of the range favored by our fit, $\gamma \sim 1.6-1.8$. 

In each frame of Fig.~\ref{many_gammas_511}, we have selected values of $m_{\rm BH}$ and $f_{\rm DM}$ which provide the best possible normalization to the INTEGRAL data. The orange bands shown in Fig.~\ref{18gamma} represent the range of these parameters that lead to the best-fit normalization for the 511 keV signal (For results using other values of $\gamma$, or for non-monochromatic mass distributions, see the Appendix.).

\begin{figure}
\includegraphics[width=3.4in,angle=0]{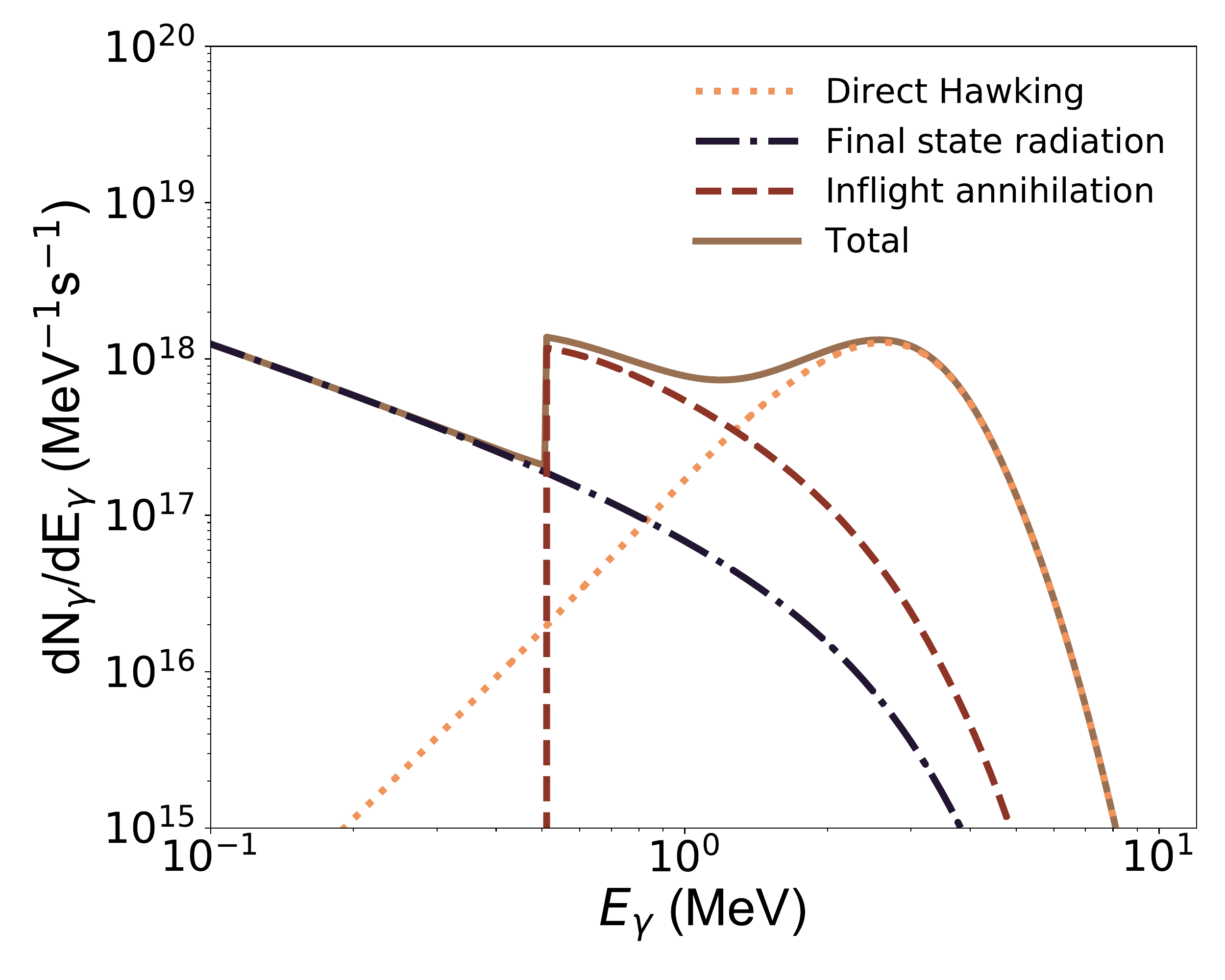}
\caption{The gamma-ray emission from a black hole with a mass of $m_{\rm BH} = 2 \times 10^{16}$\,g, showing separately the contributions from direct Hawking radiation, inflight positron annihilation, and final state radiation.}
\label{all_three}
\end{figure}

In addition to electrons and positrons, black holes produce gamma rays which can be used to constrain this class of scenarios. Photons can be produced directly as the products of Hawking evaporation (see Eq.~\ref{page}), as well as through the inflight annihilation of positrons, and as final state radiation. This inflight annihilation, final state radiation, and the contribution directly from black hole evaporation, make up our 511 keV photon emission from black holes, and takes into account electron propagation in the Milky Way. The flux of these gamma rays from a population of black holes observed over a solid angle, $\Delta \Omega$, is given by:
\begin{align} \label{jfactor}
F_{\gamma}(\Delta \Omega) &= \frac{dN^{\rm tot}_{\gamma}}{dE_{\gamma}}\frac{1}{4\pi} \int_{\Delta \Omega} \int_{los} n_{\textrm{BH}}(l, \Omega) \,dl \, d\Omega, \\
&= \frac{dN^{\rm tot}_{\gamma}}{dE_{\gamma}}\frac{f_{\rm BH}}{4\pi m_{\rm BH}} \int_{\Delta \Omega} \int_{los} \rho_{\textrm{DM}}(l, \Omega) \,dl \, d\Omega, \nonumber
\end{align}
where $dN^{\rm tot}_{\gamma}/dE_{\gamma}$ is the gamma-ray spectrum from an individual black hole, including the contributions from direct Hawking evaporation, inflight positron annihilation, and final state radiation. In Fig.~\ref{all_three}, we plot the spectrum of the gamma-ray emission from a black hole with a mass of $m_{\rm BH} = 2 \times 10^{16}$\,g, showing separately each of these contributions. At the highest energies, the direct Hawking radiation dominates, while inflight annihilation provides the largest contribution at lower energies. For the calculation of the contributions from inflight annihilation and final state radiation, see the Appendix.

To constrain the abundance of primordial black holes in the Inner Galaxy, we make use of gamma-ray measurements from the INTEGRAL~\cite{Bouchet:2011fn} and COMPTEL~\cite{Strong:1998ck} satellites (for lower mass black holes, one would instead derive constraints from gamma-ray measurements at higher energies, such as those from EGRET~\cite{Lehoucq:2009ge} or Fermi~\cite{Fermi-LAT:2018pfs}). We utilize the gamma-ray fluxes as measured in the 0.1-0.2 MeV, 0.2-0.6 MeV and 0.6-1.8 MeV bands by INTEGRAL, and in the 1-3 MeV, 3-10 MeV, and 10-30 MeV bands by COMPTEL. For each INTEGRAL band, we integrate the predicted gamma-ray flux over each of the angular bins presented. We then vary $f_{\rm BH}$ until we identify the value at which the total $\chi^2$ has increased from its best fit at the 2$\sigma$ level (for similar analyses, see Refs.~\cite{Laha:2020ivk,Cirelli:2020bpc}). For our comparison to the COMPTEL data, the large systematic (and highly correlated) error bars make it inappropriate to conduct a $\chi^2$ analysis, so we simply require that our predicted flux does not exceed any of the measurements by more than $2\sigma$ (see also, Ref.~\cite{Coogan:2020tuf}). We also apply constraints derived from the local electron/positron flux as measured by Voyager 1~\cite{Boudaud:2018hqb}.

In Fig.~\ref{18gamma}, we plot the INTEGRAL, COMPTEL, and Voyager 1 constraints on the value of $f_{\rm BH}$, as a function of $m_{\rm BH}$, for $\gamma =1.6$ or 1.8. After taking these constraints into account, we conclude that $\sim (1-4) \times 10^{16}$ g black holes could produce the observed 511 keV excess if they constitute a small fraction of the total dark matter halo, $f_{\rm BH} \sim 0.0001-0.004$, and are distributed according to a halo profile with a very steep inner slope. For this range of parameters, there are approximately $\sim 10^{22}-10^{23}$ black holes in the innermost kpc of the Milky Way. It is plausible that black holes in this mass range could have been formed in the very early universe. In particular, the horizon enclosed a total mass of $\sim 10^{16}$ g when the universe was at a temperature of $\sim10^{8}$ GeV~\cite{Hawking:1982ga, GarciaBellido:1996qt, Kawasaki:2016pql, Clesse:2016vqa, Kannike:2017bxn, Kawasaki:1997ju, Cai:2018rqf, Yoo:2018kvb, Young:2015kda, Clesse:2015wea, Hsu:1990fg, La:1989za, La:1989st, La:1989pn, PhysRevD.40.3950, Steinhardt:1990zx, Holman:1990wq, Khlopov:1980mg}. Note that although we have adopted a monochromatic mass distribution in our calculations, a distributions of masses with a variance as large as roughly an order of magnitude could plausibly produce the observed 511 keV signal without conflicting with existing constraints. We also note that in the presence of a sizable population of primordial black holes, the remaining dark matter cannot consist of particles that are capable of producing photons or other readily detectable particles through their self-annihilations~\cite{Carr:2020mqm, Adamek:2019gns}.


In the regions of parameter space in which primordial black holes could generate the observed 511 keV excess, the number density of black holes in the vicinity of the Solar System is given by:
\begin{align}
n^{\rm local}_{\rm BH} &= \frac{f_{\rm BH} \rho^{\rm local}_{\rm DM}}{m_{\rm BH}} \\
&\simeq 1.0 \times 10^{12} \, {\rm pc}^{-3} \times \bigg(\frac{f_{\rm BH}}{10^{-3}} \bigg) \, \bigg(\frac{2\times 10^{16} \, {\rm g}}{m_{\rm BH}}\bigg) \nonumber \\
&\simeq 1.2 \times 10^{-4} \, {\rm AU}^{-3} \times \bigg(\frac{f_{\rm BH}}{10^{-3}} \bigg) \, \bigg(\frac{2\times 10^{16} \, {\rm g}}{m_{\rm BH}}\bigg), \nonumber
\end{align}
where we have again adopted a local dark matter density of $\rho^{\rm local}_{\rm DM} = 0.4 \, {\rm GeV}/{\rm cm}^3$. For this number density, we should expect the closest black hole to be located at a distance of only $d \sim (3/4\pi n_{\rm BH})^{1/3} \sim \mathcal{O}(10 \, {\rm AU})$, and for the Solar System to contain several hundred black holes at any given moment. Considering the significant velocities expected of any black holes that are part of the Milky Way's dark matter halo, we conclude that black holes should regularly pass through the inner Solar System in this class of scenarios. Over a ten year window, for example, and adopting a representative velocity of 300 km/s, we predict a closest approach of a few AU or less. At such close proximity, one might think that the Hawking radiation from an individual black hole could be detectable, especially in light of proposed satellite-based gamma-ray telescopes optimized for sensitivity to MeV-scale photons, such as AMEGO~\cite{McEnery:2019tcm} and e-ASTROGAM~\cite{DeAngelis:2016slk} (see, for example, Ref.~\cite{Sobrinho:2014vka}). Even for these future missions, however, the design sensitivity to point sources of MeV-scale photons is only $\sim$\,$10^{-6} \, {\rm MeV} \, {\rm cm}^{-2} \,{\rm s}^{-1}$. In comparison, the gamma-ray flux from an individual black hole is predicted to be $\sim$\,$10^{-10}  \, {\rm MeV} \, {\rm cm}^{-2} \,{\rm s}^{-1} \times (10 \, {\rm AU}/d)^2 \, (2 \times 10^{16} \, {\rm g}/m_{\rm BH})^2$, leading to a detectable signal only for a black hole closer $d \lsim 1 {\rm AU}$. Even if we were to be lucky enough to have a black hole at such a close proximity, its proper motion would significantly complicate the search for its Hawking radiation.

A more promising way to test this class of scenarios is to use a telescope such as AMEGO~\cite{McEnery:2019tcm} or \mbox{e-ASTROGAM}~\cite{DeAngelis:2016slk} to detect and characterize the diffuse gamma-ray emission generated by black holes in the Milky Way's inner halo, or as they contribute to the isotropic gamma-ray background~\cite{Ray:2021mxu}. Instruments such as AMEGO or e-ASTROGAM should be able to improve substantially (by approximately an order of magnitude) on the limits derived in study using COMPTEL and INTEGRAL data, making it possible to definitively test the range of parameter space in which primordial black holes could produce the observed 511 keV signal.

In summary, we have shown in this letter that a population of primordial black holes could potentially produce the excess of 511 keV photons observed from the Inner Galaxy by the INTEGRAL satellite. To reproduce the angular distribution of this signal, the black holes must be distributed with a very cuspy halo profile, featuring an inner slope of $\gamma \sim 1.6$ or higher. Furthermore, to be consistent with gamma-ray constraints from COMPTEL and INTEGRAL, and cosmic-ray electron constraints from Voyager 1, the bulk of these black holes must have masses in the range of $m_{\rm BH} \sim (1-4)\times 10^{16} \, {\rm g}$. To provide the observed normalization of this signal, these black holes must constitute a relatively small fraction of the total dark matter abundance, $f_{\rm DM} \sim 0.0001-0.004$.

In our calculations, we have included the photons, electrons and positrons that are directly produced through Hawking radiation, as well as the gamma-rays that are generated through the inflight annihilation of positrons, and as final state radiation. For the range of $m_{\rm BH}$ and $f_{\rm DM}$ required in this scenario, we predict that the local halo of the Milky Way should contain a considerable number density of black holes, $n^{\rm local}_{\rm BH} \sim 10^{12} \, {\rm pc}^{-3}$. Although such black holes would be very challenging to detect individually, future satellite-based gamma-ray telescopes such as AMEGO or e-ASTROGAM will be able to definitively test this class of scenarios by measuring the spectrum and morphology of the MeV-scale emission from the Inner Galaxy.

  \bigskip

\begin{acknowledgments}  
CK would like to thank Cory Cotter for helpful discussions. CK is supported by the National Science Foundation Graduate Research Fellowship Program under Grant No. DGE-1746045. DH is supported by the Fermi Research Alliance, LLC under Contract No. DE-AC02-07CH11359 with the U.S. Department of Energy, Office of Science, Office of High Energy Physics. 

\end{acknowledgments}

\clearpage

\onecolumngrid

\twocolumngrid

\onecolumngrid

\appendix

\section{Inflight Annihilation }
\label{supplementarymaterial1}

In the energy range of interest for this study, most positrons lose energy via ionization and become non-relativistic before annihilating. A small fraction of such positrons, however, will annihilate inflight prior to reaching non-relativistic velocities. Such annihilations can produce photons with energies greater than 511 keV, thus contributing to the continuum spectrum of diffuse gamma rays. 

Following Ref.~\cite{Beacom:2004pe}, the spectrum of gamma rays from the inflight annihilation of positrons is given by:  
%
\begin{align}
\label{eq:IA}
\frac{dN^{\rm IA}_{\gamma}}{dE_{\gamma}} &= \frac{\pi \alpha^2 n_H}{m_e}  \int^{\infty}_{m_e} dE_{e^+} \frac{dN_{e^+}}{dE_{e^+}} \int^{E_{e^+}}_{E_{\rm min}} \frac{dE}{dE/dx} \, \frac{P_{E_{e^+} \rightarrow E} }{(E^2-m_e^2)} \\ 
&\times  \bigg(-2-\frac{(E+m_e)(m_e^2(E+m_e)+E_{\gamma}^2(E+3m_e)-E_{\gamma}(E+m_e)(E+3m_e))}{E_{\gamma}^2(E-E_{\gamma}+m_e)^2} \nonumber
\bigg), \nonumber 
\end{align}
%
where $\alpha$ is the fine structure constant, $n_H$ is the number density of neutral hydrogen atoms, and $dN_{e^+}/dE_{e^+}$ is the spectrum of positrons radiated from the black hole. $dE/dx$ is the energy loss rate of a positron due to ionization in the presence of neutral hydrogen, as described by the Bethe-Bloch formula. Since $dE/dx$ is proportional to $n_H$, the flux of gamma rays from inflight annihilation is not sensitive to the value of the gas density.

The probability that a positron of energy $E_{e^+}$ will survive until its energy has been reduced to $E$ is given by
\begin{align}
P_{E_{e^+} \rightarrow E} = \exp\bigg(-n_H \int^{E_{e^+}}_E \sigma_{\rm ann}(E') \,\frac{dE'}{|dE'/dx|}  \bigg),
\end{align}
where $\sigma_{\rm ann}$ is the cross section for a positron annihilating with an electron at rest. For positrons with an energy below a few MeV, $P_{E_{e^+} \rightarrow m_e}$ is always between $\sim$\,0.95 and unity. In other words, only a few percent of the positrons annihilate before becoming non-relativistic.

\section{Final State Radiation}

Electrons, positrons, and other charged particles emitted from a black hole can radiate photons as final state radiation. This process contributes as follows to the gamma-ray spectrum:
\begin{align}
\frac{dN^{\rm FSR}_{\gamma}}{dE_{\gamma}} &=\frac{\alpha}{2\pi} \int dE_{e} \frac{dN_e}{dE_e} \,  \bigg( \frac{2}{E_{\gamma}} +\frac{E_{\gamma}}{E^2_e} - \frac{2}{E_e} \bigg) \, \bigg[\ln\bigg(\frac{2E_e(E_e-E_{\gamma})}{m^2_e}\bigg)-1\bigg], 
\end{align}
where $dN_{e}/dE_{e}$ is the spectrum of electrons and positrons radiated from the black hole.

\begin{figure*}
\includegraphics[width=3.4in,angle=0]{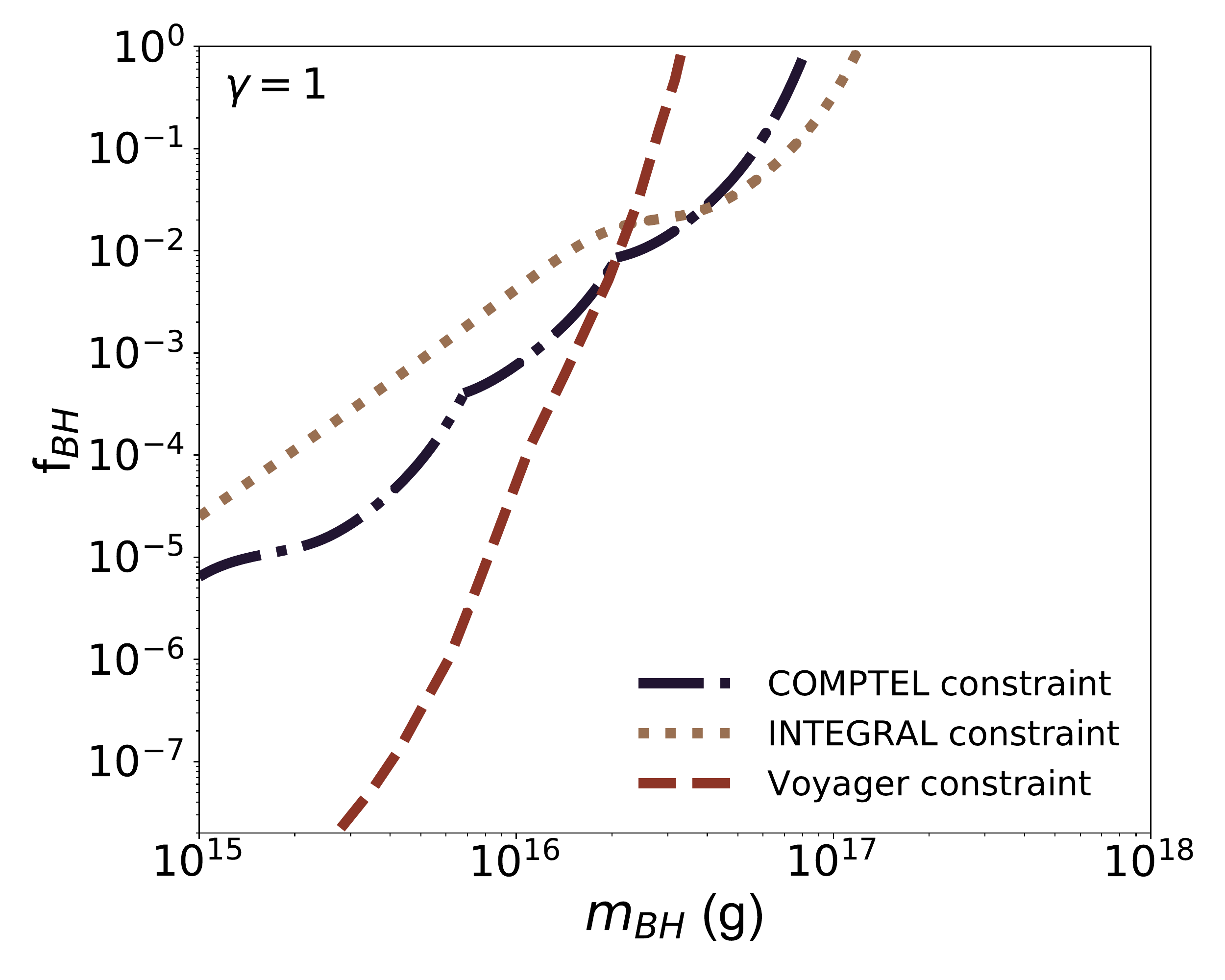}
\includegraphics[width=3.4in,angle=0]{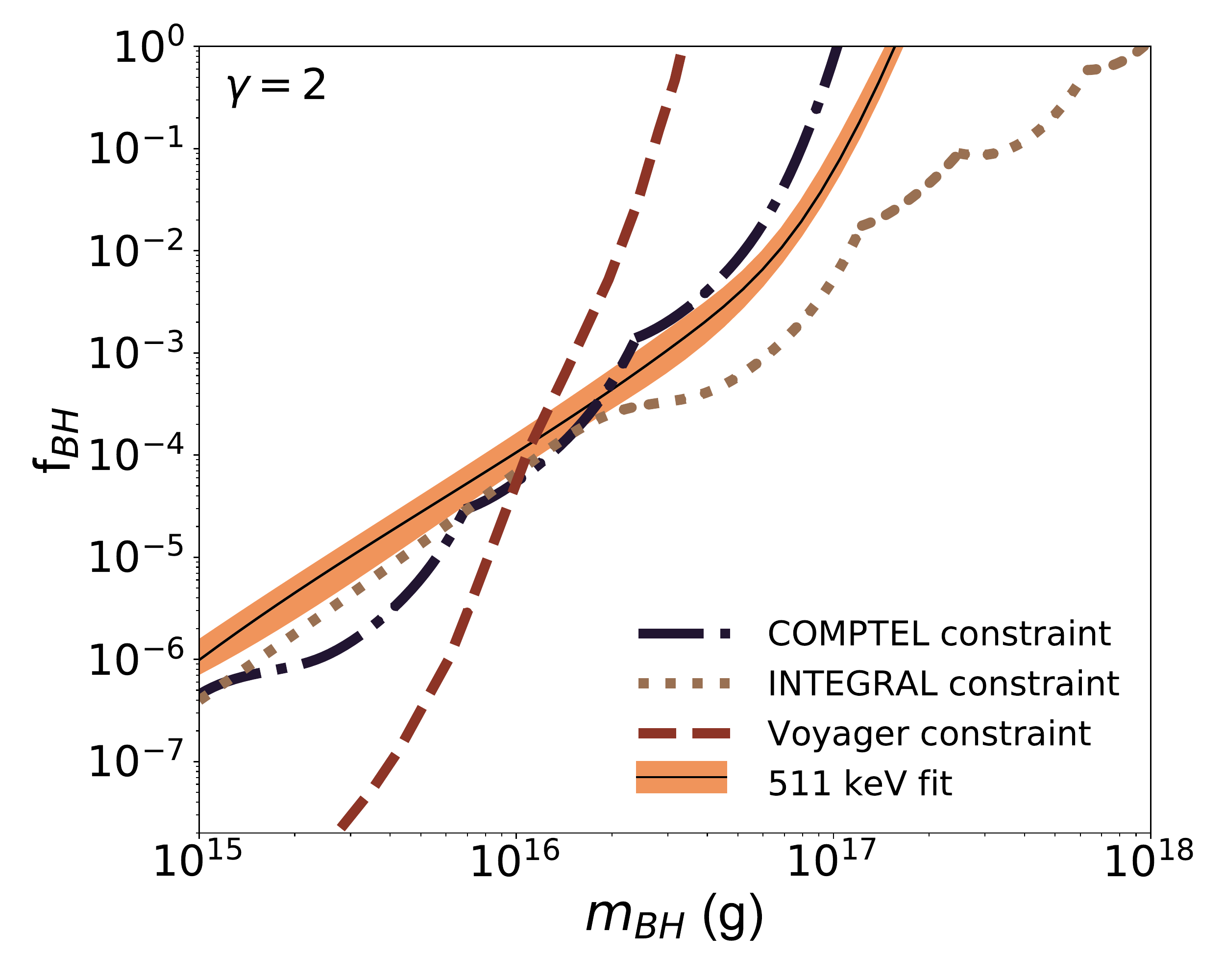}
\caption{As in Fig.~2, but for black hole distributions with an inner slope of $\gamma=1$ (left) or $\gamma=2$ (right). In the $\gamma=1$ case, we do not show any region for the 511 keV excess, as the angular distribution of this signal cannot be accommodated for this choice of halo profile. In the $\gamma=2$ case, the region favored by the 511 keV excess is ruled out by a combination of the constraints from COMPTEL, INTEGRAL, and Voyager 1.}
\label{1gamma}
\end{figure*}

\section{Results For Other Values of $\gamma$}

In Fig.~2, we showed the regions of black hole parameter space that can accommodate the observed characteristics of the 511 keV excess, along with the constraints from gamma-ray and cosmic-ray electron measurements, for halo profiles with an inner slope of $\gamma=1.6$ or 1.8. In Fig.~\ref{1gamma}, we show the corresponding results for the cases of $\gamma=1$ or $\gamma=2$. In the $\gamma=1$ frame, we do not show any region for the 511 keV excess, as the angular distribution of this signal cannot be accommodated for this choice of halo profile. In the $\gamma=2$ case, the region favored by the 511 keV excess is ruled out by a combination of the constraints from COMPTEL, INTEGRAL, and Voyager 1.

\begin{figure}
\includegraphics[width=3.4in,angle=0]{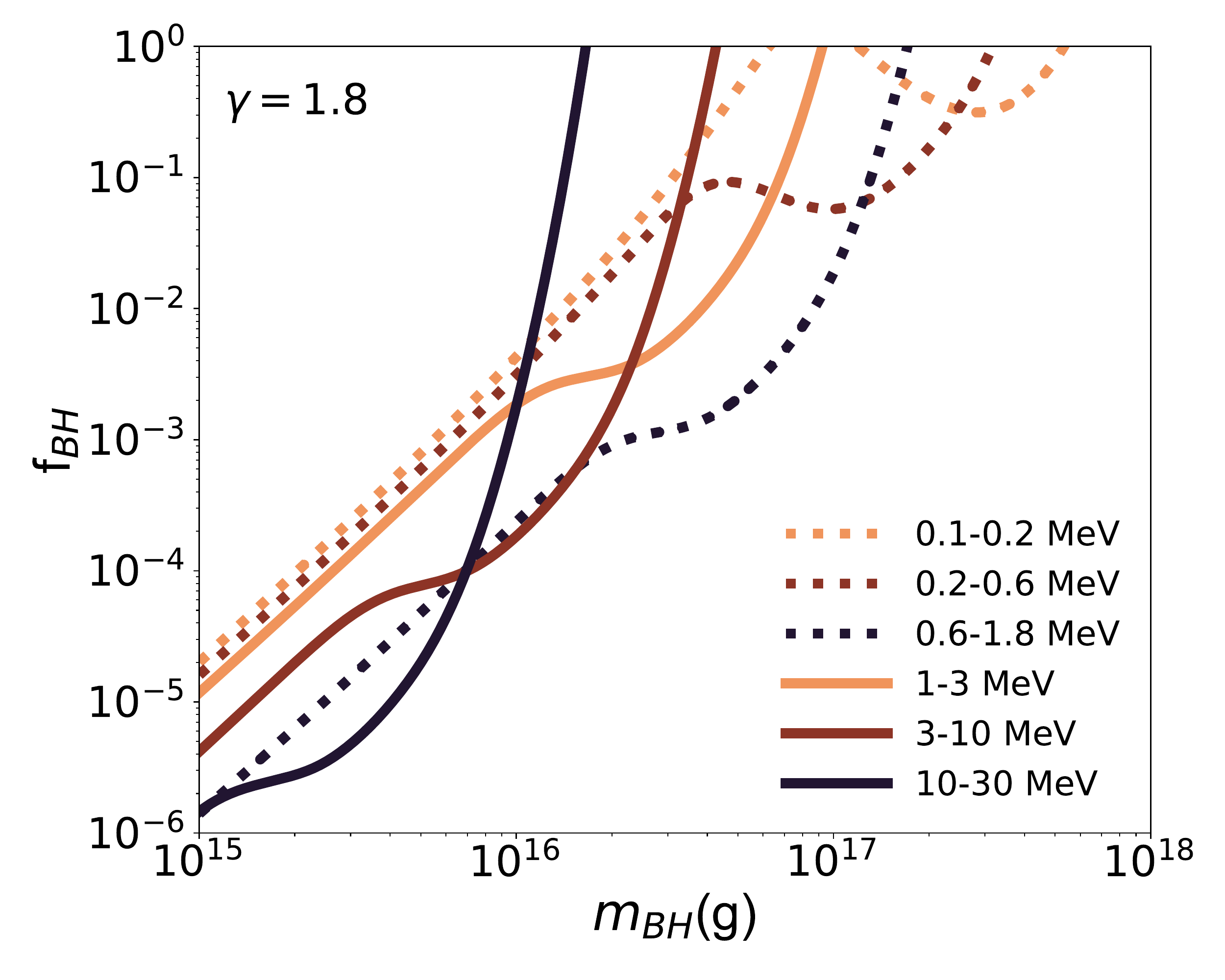}
\caption{the constraints on the black hole parameter space from individual energy bins of data from the INTEGRAL (dashed) and COMPTEL (solid) satellites, for the case of $\gamma=1.8$.}
\label{allcons}
\end{figure}

\section{Bin-by-Bin Gamma-Ray Constraints}

In Fig.~\ref{allcons}, we show the constraints on the black hole parameter space as derived from individual energy bins of INTEGRAL and COMPTEL data, for the case of $\gamma=1.8$. INTEGRAL provides the most stringent constraints on black holes more massive than $\sim 2 \times 10^{16} \, {\rm g}$, while COMPTEL (and Voyager 1) are most restrictive for smaller values of $m_{\rm BH}$.

\section{Non-Monocromatic Mass Spectrum}
\begin{figure*}
\includegraphics[width=3.4in,angle=0]{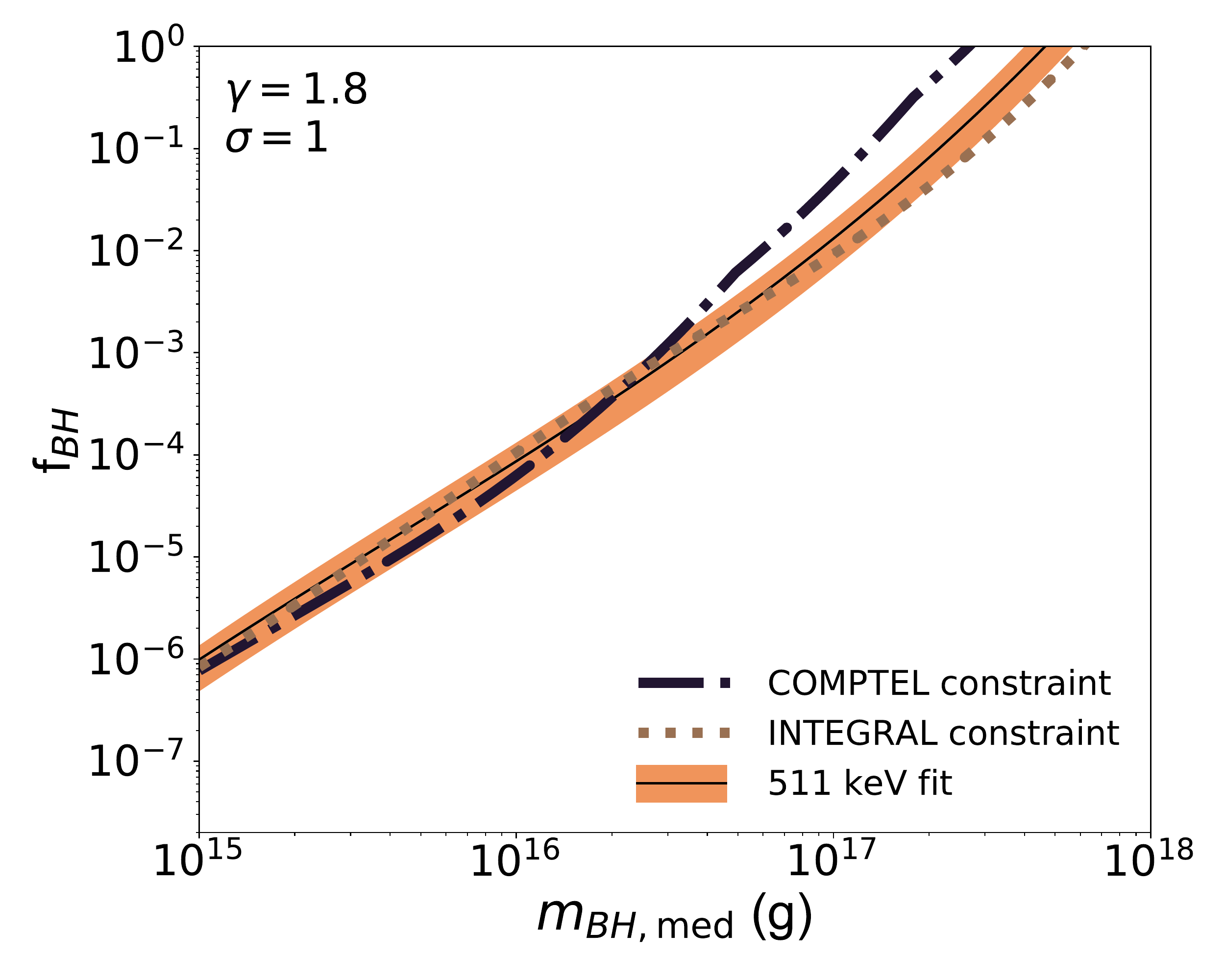}
\includegraphics[width=3.4in,angle=0]{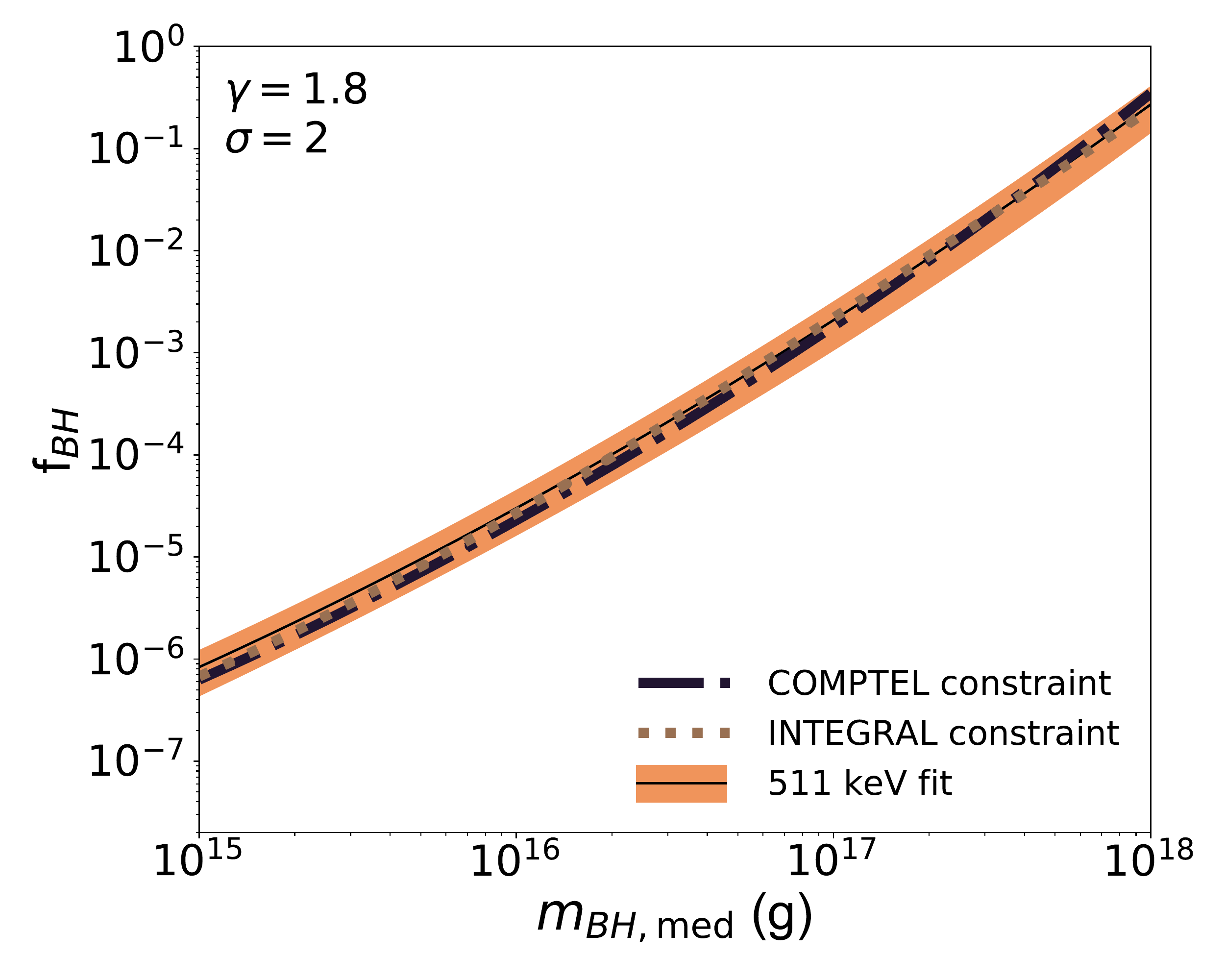}
\caption{As in Fig.~2, but for black hole distributions with an inner slope of $\gamma=1.8$, with a non-monochromatic mass distribution. We do not show the Voyager constraints, as they are specific to a monochromatic mass spectrum. The non-monochromatic mass spectrum is calculated using a log normal distribution, with the median black hole mass being plotted on the $x$ axis, with $\sigma=1$ in the left plot, and $\sigma=2$ on the right.}
\label{nonmono}
\end{figure*}

In addition, we show the constraints in the black hole parameter space for a non-monochromatic mass spectrum in Fig.~\ref{nonmono}. For this mass spectrum, we follow a log-normal distribution and plot the parameter space as a function of median black hole mass, with $\sigma=1$ (left) and $\sigma=2$ (right). We consider a value of $\gamma=1.8$. For a Gaussian distribution of black holes masses, the constraints from COMPTEL and/or INTEGRAL are approximately saturated across a wide range of median mass values. In such a scenario, a large fraction of the gamma-ray emission observed from the Inner Galaxy must originate from PBHs.

\bibliography{511excess_celestev6}

\begin{thebibliography}{104}%
\makeatletter
\providecommand \@ifxundefined [1]{%
 \@ifx{#1\undefined}
}%
\providecommand \@ifnum [1]{%
 \ifnum #1\expandafter \@firstoftwo
 \else \expandafter \@secondoftwo
 \fi
}%
\providecommand \@ifx [1]{%
 \ifx #1\expandafter \@firstoftwo
 \else \expandafter \@secondoftwo
 \fi
}%
\providecommand \natexlab [1]{#1}%
\providecommand \enquote  [1]{``#1''}%
\providecommand \bibnamefont  [1]{#1}%
\providecommand \bibfnamefont [1]{#1}%
\providecommand \citenamefont [1]{#1}%
\providecommand \href@noop [0]{\@secondoftwo}%
\providecommand \href [0]{\begingroup \@sanitize@url \@href}%
\providecommand \@href[1]{\@@startlink{#1}\@@href}%
\providecommand \@@href[1]{\endgroup#1\@@endlink}%
\providecommand \@sanitize@url [0]{\catcode `\\12\catcode `\$12\catcode
  `\&12\catcode `\#12\catcode `\^12\catcode `\_12\catcode `\%12\relax}%
\providecommand \@@startlink[1]{}%
\providecommand \@@endlink[0]{}%
\providecommand \url  [0]{\begingroup\@sanitize@url \@url }%
\providecommand \@url [1]{\endgroup\@href {#1}{\urlprefix }}%
\providecommand \urlprefix  [0]{URL }%
\providecommand \Eprint [0]{\href }%
\providecommand \doibase [0]{http://dx.doi.org/}%
\providecommand \selectlanguage [0]{\@gobble}%
\providecommand \bibinfo  [0]{\@secondoftwo}%
\providecommand \bibfield  [0]{\@secondoftwo}%
\providecommand \translation [1]{[#1]}%
\providecommand \BibitemOpen [0]{}%
\providecommand \bibitemStop [0]{}%
\providecommand \bibitemNoStop [0]{.\EOS\space}%
\providecommand \EOS [0]{\spacefactor3000\relax}%
\providecommand \BibitemShut  [1]{\csname bibitem#1\endcsname}%
\let\auto@bib@innerbib\@empty
\bibitem [{\citenamefont {Weidenspointner}\ \emph {et~al.}(2004)\citenamefont
  {Weidenspointner} \emph {et~al.}}]{Weidenspointner:2004my}%
  \BibitemOpen
  \bibfield  {author} {\bibinfo {author} {\bibfnamefont {G.}~\bibnamefont
  {Weidenspointner}} \emph {et~al.},\ }in\ \href@noop {} {\emph {\bibinfo
  {booktitle} {{5th INTEGRAL Workshop}: {The INTEGRAL Universe}}}}\ (\bibinfo
  {year} {2004})\ \Eprint {http://arxiv.org/abs/astro-ph/0406178}
  {arXiv:astro-ph/0406178} \BibitemShut {NoStop}%
\bibitem [{\citenamefont {Churazov}\ \emph {et~al.}(2005)\citenamefont
  {Churazov}, \citenamefont {Sunyaev}, \citenamefont {Sazonov}, \citenamefont
  {Revnivtsev},\ and\ \citenamefont {Varshalovich}}]{Churazov:2004as}%
  \BibitemOpen
  \bibfield  {author} {\bibinfo {author} {\bibfnamefont {E.}~\bibnamefont
  {Churazov}}, \bibinfo {author} {\bibfnamefont {R.}~\bibnamefont {Sunyaev}},
  \bibinfo {author} {\bibfnamefont {S.}~\bibnamefont {Sazonov}}, \bibinfo
  {author} {\bibfnamefont {M.}~\bibnamefont {Revnivtsev}}, \ and\ \bibinfo
  {author} {\bibfnamefont {D.}~\bibnamefont {Varshalovich}},\ }\href {\doibase
  10.1111/j.1365-2966.2005.08757.x} {\bibfield  {journal} {\bibinfo  {journal}
  {Mon. Not. Roy. Astron. Soc.}\ }\textbf {\bibinfo {volume} {357}},\ \bibinfo
  {pages} {1377} (\bibinfo {year} {2005})},\ \Eprint
  {http://arxiv.org/abs/astro-ph/0411351} {arXiv:astro-ph/0411351} \BibitemShut
  {NoStop}%
\bibitem [{\citenamefont {Weidenspointner}\ \emph {et~al.}(2007)\citenamefont
  {Weidenspointner} \emph {et~al.}}]{Weidenspointner:2007rs}%
  \BibitemOpen
  \bibfield  {author} {\bibinfo {author} {\bibfnamefont {G.}~\bibnamefont
  {Weidenspointner}} \emph {et~al.},\ }\href@noop {} {\bibfield  {journal}
  {\bibinfo  {journal} {ESA Spec. Publ.}\ }\textbf {\bibinfo {volume} {622}},\
  \bibinfo {pages} {25} (\bibinfo {year} {2007})},\ \Eprint
  {http://arxiv.org/abs/astro-ph/0702621} {arXiv:astro-ph/0702621} \BibitemShut
  {NoStop}%
\bibitem [{\citenamefont {Jean}\ \emph {et~al.}(2006)\citenamefont {Jean},
  \citenamefont {Knodlseder}, \citenamefont {Gillard}, \citenamefont
  {Guessoum}, \citenamefont {Ferriere}, \citenamefont {Marcowith},
  \citenamefont {Lonjou},\ and\ \citenamefont {Roques}}]{Jean:2005af}%
  \BibitemOpen
  \bibfield  {author} {\bibinfo {author} {\bibfnamefont {P.}~\bibnamefont
  {Jean}}, \bibinfo {author} {\bibfnamefont {J.}~\bibnamefont {Knodlseder}},
  \bibinfo {author} {\bibfnamefont {W.}~\bibnamefont {Gillard}}, \bibinfo
  {author} {\bibfnamefont {N.}~\bibnamefont {Guessoum}}, \bibinfo {author}
  {\bibfnamefont {K.}~\bibnamefont {Ferriere}}, \bibinfo {author}
  {\bibfnamefont {A.}~\bibnamefont {Marcowith}}, \bibinfo {author}
  {\bibfnamefont {V.}~\bibnamefont {Lonjou}}, \ and\ \bibinfo {author}
  {\bibfnamefont {J.~P.}\ \bibnamefont {Roques}},\ }\href {\doibase
  10.1051/0004-6361:20053765} {\bibfield  {journal} {\bibinfo  {journal}
  {Astron. Astrophys.}\ }\textbf {\bibinfo {volume} {445}},\ \bibinfo {pages}
  {579} (\bibinfo {year} {2006})},\ \Eprint
  {http://arxiv.org/abs/astro-ph/0509298} {arXiv:astro-ph/0509298} \BibitemShut
  {NoStop}%
\bibitem [{\citenamefont {Weidenspointner}\ \emph {et~al.}(2008)\citenamefont
  {Weidenspointner} \emph {et~al.}}]{Weidenspointner:2008zz}%
  \BibitemOpen
  \bibfield  {author} {\bibinfo {author} {\bibfnamefont {G.}~\bibnamefont
  {Weidenspointner}} \emph {et~al.},\ }\href {\doibase 10.1038/nature06490}
  {\bibfield  {journal} {\bibinfo  {journal} {Nature}\ }\textbf {\bibinfo
  {volume} {451}},\ \bibinfo {pages} {159} (\bibinfo {year}
  {2008})}\BibitemShut {NoStop}%
\bibitem [{\citenamefont {Kierans}\ \emph {et~al.}(2020)\citenamefont {Kierans}
  \emph {et~al.}}]{Kierans:2019aqz}%
  \BibitemOpen
  \bibfield  {author} {\bibinfo {author} {\bibfnamefont {C.~A.}\ \bibnamefont
  {Kierans}} \emph {et~al.},\ }\href {\doibase 10.3847/1538-4357/ab89a9}
  {\bibfield  {journal} {\bibinfo  {journal} {Astrophys. J.}\ }\textbf
  {\bibinfo {volume} {895}},\ \bibinfo {pages} {44} (\bibinfo {year} {2020})},\
  \Eprint {http://arxiv.org/abs/1912.00110} {arXiv:1912.00110 [astro-ph.HE]}
  \BibitemShut {NoStop}%
\bibitem [{\citenamefont {Prantzos}(2006)}]{Prantzos:2005pz}%
  \BibitemOpen
  \bibfield  {author} {\bibinfo {author} {\bibfnamefont {N.}~\bibnamefont
  {Prantzos}},\ }\href {\doibase 10.1051/0004-6361:20052811} {\bibfield
  {journal} {\bibinfo  {journal} {Astron. Astrophys.}\ }\textbf {\bibinfo
  {volume} {449}},\ \bibinfo {pages} {869} (\bibinfo {year} {2006})},\ \Eprint
  {http://arxiv.org/abs/astro-ph/0511190} {arXiv:astro-ph/0511190} \BibitemShut
  {NoStop}%
\bibitem [{\citenamefont {Kalemci}\ \emph {et~al.}(2006)\citenamefont
  {Kalemci}, \citenamefont {Boggs}, \citenamefont {Milne},\ and\ \citenamefont
  {Reynolds}}]{Kalemci:2006bz}%
  \BibitemOpen
  \bibfield  {author} {\bibinfo {author} {\bibfnamefont {E.}~\bibnamefont
  {Kalemci}}, \bibinfo {author} {\bibfnamefont {S.~E.}\ \bibnamefont {Boggs}},
  \bibinfo {author} {\bibfnamefont {P.~A.}\ \bibnamefont {Milne}}, \ and\
  \bibinfo {author} {\bibfnamefont {S.~P.}\ \bibnamefont {Reynolds}},\ }\href
  {\doibase 10.1086/503289} {\bibfield  {journal} {\bibinfo  {journal}
  {Astrophys. J. Lett.}\ }\textbf {\bibinfo {volume} {640}},\ \bibinfo {pages}
  {L55} (\bibinfo {year} {2006})},\ \Eprint
  {http://arxiv.org/abs/astro-ph/0602233} {arXiv:astro-ph/0602233} \BibitemShut
  {NoStop}%
\bibitem [{\citenamefont {Casse}\ \emph {et~al.}(2004)\citenamefont {Casse},
  \citenamefont {Cordier}, \citenamefont {Paul},\ and\ \citenamefont
  {Schanne}}]{Casse:2003fh}%
  \BibitemOpen
  \bibfield  {author} {\bibinfo {author} {\bibfnamefont {M.}~\bibnamefont
  {Casse}}, \bibinfo {author} {\bibfnamefont {B.}~\bibnamefont {Cordier}},
  \bibinfo {author} {\bibfnamefont {J.}~\bibnamefont {Paul}}, \ and\ \bibinfo
  {author} {\bibfnamefont {S.}~\bibnamefont {Schanne}},\ }\href {\doibase
  10.1086/381884} {\bibfield  {journal} {\bibinfo  {journal} {Astrophys. J.
  Lett.}\ }\textbf {\bibinfo {volume} {602}},\ \bibinfo {pages} {L17} (\bibinfo
  {year} {2004})},\ \Eprint {http://arxiv.org/abs/astro-ph/0309824}
  {arXiv:astro-ph/0309824} \BibitemShut {NoStop}%
\bibitem [{\citenamefont {Bertone}\ \emph {et~al.}(2006)\citenamefont
  {Bertone}, \citenamefont {Kusenko}, \citenamefont {Palomares-Ruiz},
  \citenamefont {Pascoli},\ and\ \citenamefont {Semikoz}}]{Bertone:2004ek}%
  \BibitemOpen
  \bibfield  {author} {\bibinfo {author} {\bibfnamefont {G.}~\bibnamefont
  {Bertone}}, \bibinfo {author} {\bibfnamefont {A.}~\bibnamefont {Kusenko}},
  \bibinfo {author} {\bibfnamefont {S.}~\bibnamefont {Palomares-Ruiz}},
  \bibinfo {author} {\bibfnamefont {S.}~\bibnamefont {Pascoli}}, \ and\
  \bibinfo {author} {\bibfnamefont {D.}~\bibnamefont {Semikoz}},\ }\href
  {\doibase 10.1016/j.physletb.2006.03.022} {\bibfield  {journal} {\bibinfo
  {journal} {Phys. Lett. B}\ }\textbf {\bibinfo {volume} {636}},\ \bibinfo
  {pages} {20} (\bibinfo {year} {2006})},\ \Eprint
  {http://arxiv.org/abs/astro-ph/0405005} {arXiv:astro-ph/0405005} \BibitemShut
  {NoStop}%
\bibitem [{\citenamefont {Guessoum}\ \emph {et~al.}(2006)\citenamefont
  {Guessoum}, \citenamefont {Jean},\ and\ \citenamefont
  {Prantzos}}]{Guessoum:2006fs}%
  \BibitemOpen
  \bibfield  {author} {\bibinfo {author} {\bibfnamefont {N.}~\bibnamefont
  {Guessoum}}, \bibinfo {author} {\bibfnamefont {P.}~\bibnamefont {Jean}}, \
  and\ \bibinfo {author} {\bibfnamefont {N.}~\bibnamefont {Prantzos}},\ }\href
  {\doibase 10.1051/0004-6361:20065240} {\bibfield  {journal} {\bibinfo
  {journal} {Astron. Astrophys.}\ }\textbf {\bibinfo {volume} {457}},\ \bibinfo
  {pages} {753} (\bibinfo {year} {2006})},\ \Eprint
  {http://arxiv.org/abs/astro-ph/0607296} {arXiv:astro-ph/0607296} \BibitemShut
  {NoStop}%
\bibitem [{\citenamefont {Bartels}\ \emph {et~al.}(2018)\citenamefont
  {Bartels}, \citenamefont {Calore}, \citenamefont {Storm},\ and\ \citenamefont
  {Weniger}}]{Bartels:2018eyb}%
  \BibitemOpen
  \bibfield  {author} {\bibinfo {author} {\bibfnamefont {R.}~\bibnamefont
  {Bartels}}, \bibinfo {author} {\bibfnamefont {F.}~\bibnamefont {Calore}},
  \bibinfo {author} {\bibfnamefont {E.}~\bibnamefont {Storm}}, \ and\ \bibinfo
  {author} {\bibfnamefont {C.}~\bibnamefont {Weniger}},\ }\href {\doibase
  10.1093/mnras/sty2135} {\bibfield  {journal} {\bibinfo  {journal} {Mon. Not.
  Roy. Astron. Soc.}\ }\textbf {\bibinfo {volume} {480}},\ \bibinfo {pages}
  {3826} (\bibinfo {year} {2018})},\ \Eprint {http://arxiv.org/abs/1803.04370}
  {arXiv:1803.04370 [astro-ph.HE]} \BibitemShut {NoStop}%
\bibitem [{\citenamefont {Takhistov}(2020)}]{Takhistov:2019zyb}%
  \BibitemOpen
  \bibfield  {author} {\bibinfo {author} {\bibfnamefont {V.}~\bibnamefont
  {Takhistov}},\ }\href {\doibase 10.22323/1.358.0803} {\bibfield  {journal}
  {\bibinfo  {journal} {PoS}\ }\textbf {\bibinfo {volume} {ICRC2019}},\
  \bibinfo {pages} {803} (\bibinfo {year} {2020})},\ \Eprint
  {http://arxiv.org/abs/1908.01100} {arXiv:1908.01100 [astro-ph.HE]}
  \BibitemShut {NoStop}%
\bibitem [{\citenamefont {Fuller}\ \emph {et~al.}(2019)\citenamefont {Fuller},
  \citenamefont {Kusenko}, \citenamefont {Radice},\ and\ \citenamefont
  {Takhistov}}]{Fuller:2018ttb}%
  \BibitemOpen
  \bibfield  {author} {\bibinfo {author} {\bibfnamefont {G.~M.}\ \bibnamefont
  {Fuller}}, \bibinfo {author} {\bibfnamefont {A.}~\bibnamefont {Kusenko}},
  \bibinfo {author} {\bibfnamefont {D.}~\bibnamefont {Radice}}, \ and\ \bibinfo
  {author} {\bibfnamefont {V.}~\bibnamefont {Takhistov}},\ }\href {\doibase
  10.1103/PhysRevLett.122.121101} {\bibfield  {journal} {\bibinfo  {journal}
  {Phys. Rev. Lett.}\ }\textbf {\bibinfo {volume} {122}},\ \bibinfo {pages}
  {121101} (\bibinfo {year} {2019})},\ \Eprint
  {http://arxiv.org/abs/1811.00133} {arXiv:1811.00133 [astro-ph.HE]}
  \BibitemShut {NoStop}%
\bibitem [{\citenamefont {Prantzos}\ \emph {et~al.}(2011)\citenamefont
  {Prantzos} \emph {et~al.}}]{Prantzos:2010wi}%
  \BibitemOpen
  \bibfield  {author} {\bibinfo {author} {\bibfnamefont {N.}~\bibnamefont
  {Prantzos}} \emph {et~al.},\ }\href {\doibase 10.1103/RevModPhys.83.1001}
  {\bibfield  {journal} {\bibinfo  {journal} {Rev. Mod. Phys.}\ }\textbf
  {\bibinfo {volume} {83}},\ \bibinfo {pages} {1001} (\bibinfo {year}
  {2011})},\ \Eprint {http://arxiv.org/abs/1009.4620} {arXiv:1009.4620
  [astro-ph.HE]} \BibitemShut {NoStop}%
\bibitem [{\citenamefont {Boehm}\ \emph {et~al.}(2004)\citenamefont {Boehm},
  \citenamefont {Hooper}, \citenamefont {Silk}, \citenamefont {Casse},\ and\
  \citenamefont {Paul}}]{Boehm:2003bt}%
  \BibitemOpen
  \bibfield  {author} {\bibinfo {author} {\bibfnamefont {C.}~\bibnamefont
  {Boehm}}, \bibinfo {author} {\bibfnamefont {D.}~\bibnamefont {Hooper}},
  \bibinfo {author} {\bibfnamefont {J.}~\bibnamefont {Silk}}, \bibinfo {author}
  {\bibfnamefont {M.}~\bibnamefont {Casse}}, \ and\ \bibinfo {author}
  {\bibfnamefont {J.}~\bibnamefont {Paul}},\ }\href {\doibase
  10.1103/PhysRevLett.92.101301} {\bibfield  {journal} {\bibinfo  {journal}
  {Phys. Rev. Lett.}\ }\textbf {\bibinfo {volume} {92}},\ \bibinfo {pages}
  {101301} (\bibinfo {year} {2004})},\ \Eprint
  {http://arxiv.org/abs/astro-ph/0309686} {arXiv:astro-ph/0309686} \BibitemShut
  {NoStop}%
\bibitem [{\citenamefont {Huh}\ \emph {et~al.}(2008)\citenamefont {Huh},
  \citenamefont {Kim}, \citenamefont {Park},\ and\ \citenamefont
  {Park}}]{Huh:2007zw}%
  \BibitemOpen
  \bibfield  {author} {\bibinfo {author} {\bibfnamefont {J.-H.}\ \bibnamefont
  {Huh}}, \bibinfo {author} {\bibfnamefont {J.~E.}\ \bibnamefont {Kim}},
  \bibinfo {author} {\bibfnamefont {J.-C.}\ \bibnamefont {Park}}, \ and\
  \bibinfo {author} {\bibfnamefont {S.~C.}\ \bibnamefont {Park}},\ }\href
  {\doibase 10.1103/PhysRevD.77.123503} {\bibfield  {journal} {\bibinfo
  {journal} {Phys. Rev. D}\ }\textbf {\bibinfo {volume} {77}},\ \bibinfo
  {pages} {123503} (\bibinfo {year} {2008})},\ \Eprint
  {http://arxiv.org/abs/0711.3528} {arXiv:0711.3528 [astro-ph]} \BibitemShut
  {NoStop}%
\bibitem [{\citenamefont {Hooper}\ and\ \citenamefont
  {Zurek}(2008)}]{Hooper:2008im}%
  \BibitemOpen
  \bibfield  {author} {\bibinfo {author} {\bibfnamefont {D.}~\bibnamefont
  {Hooper}}\ and\ \bibinfo {author} {\bibfnamefont {K.~M.}\ \bibnamefont
  {Zurek}},\ }\href {\doibase 10.1103/PhysRevD.77.087302} {\bibfield  {journal}
  {\bibinfo  {journal} {Phys. Rev. D}\ }\textbf {\bibinfo {volume} {77}},\
  \bibinfo {pages} {087302} (\bibinfo {year} {2008})},\ \Eprint
  {http://arxiv.org/abs/0801.3686} {arXiv:0801.3686 [hep-ph]} \BibitemShut
  {NoStop}%
\bibitem [{\citenamefont {Khalil}\ and\ \citenamefont
  {Seto}(2008)}]{Khalil:2008kp}%
  \BibitemOpen
  \bibfield  {author} {\bibinfo {author} {\bibfnamefont {S.}~\bibnamefont
  {Khalil}}\ and\ \bibinfo {author} {\bibfnamefont {O.}~\bibnamefont {Seto}},\
  }\href {\doibase 10.1088/1475-7516/2008/10/024} {\bibfield  {journal}
  {\bibinfo  {journal} {JCAP}\ }\textbf {\bibinfo {volume} {10}},\ \bibinfo
  {pages} {024} (\bibinfo {year} {2008})},\ \Eprint
  {http://arxiv.org/abs/0804.0336} {arXiv:0804.0336 [hep-ph]} \BibitemShut
  {NoStop}%
\bibitem [{\citenamefont {Wilkinson}\ \emph {et~al.}(2016)\citenamefont
  {Wilkinson}, \citenamefont {Vincent}, \citenamefont {B\oe{}hm},\ and\
  \citenamefont {McCabe}}]{Wilkinson:2016gsy}%
  \BibitemOpen
  \bibfield  {author} {\bibinfo {author} {\bibfnamefont {R.~J.}\ \bibnamefont
  {Wilkinson}}, \bibinfo {author} {\bibfnamefont {A.~C.}\ \bibnamefont
  {Vincent}}, \bibinfo {author} {\bibfnamefont {C.}~\bibnamefont {B\oe{}hm}}, \
  and\ \bibinfo {author} {\bibfnamefont {C.}~\bibnamefont {McCabe}},\ }\href
  {\doibase 10.1103/PhysRevD.94.103525} {\bibfield  {journal} {\bibinfo
  {journal} {Phys. Rev. D}\ }\textbf {\bibinfo {volume} {94}},\ \bibinfo
  {pages} {103525} (\bibinfo {year} {2016})},\ \Eprint
  {http://arxiv.org/abs/1602.01114} {arXiv:1602.01114 [astro-ph.CO]}
  \BibitemShut {NoStop}%
\bibitem [{\citenamefont {Ema}\ \emph {et~al.}(2021)\citenamefont {Ema},
  \citenamefont {Sala},\ and\ \citenamefont {Sato}}]{Ema:2020fit}%
  \BibitemOpen
  \bibfield  {author} {\bibinfo {author} {\bibfnamefont {Y.}~\bibnamefont
  {Ema}}, \bibinfo {author} {\bibfnamefont {F.}~\bibnamefont {Sala}}, \ and\
  \bibinfo {author} {\bibfnamefont {R.}~\bibnamefont {Sato}},\ }\href {\doibase
  10.1140/epjc/s10052-021-08899-y} {\bibfield  {journal} {\bibinfo  {journal}
  {Eur. Phys. J. C}\ }\textbf {\bibinfo {volume} {81}},\ \bibinfo {pages} {129}
  (\bibinfo {year} {2021})},\ \Eprint {http://arxiv.org/abs/2007.09105}
  {arXiv:2007.09105 [hep-ph]} \BibitemShut {NoStop}%
\bibitem [{\citenamefont {Escudero}(2019)}]{Escudero:2018mvt}%
  \BibitemOpen
  \bibfield  {author} {\bibinfo {author} {\bibfnamefont {M.}~\bibnamefont
  {Escudero}},\ }\href {\doibase 10.1088/1475-7516/2019/02/007} {\bibfield
  {journal} {\bibinfo  {journal} {JCAP}\ }\textbf {\bibinfo {volume} {02}},\
  \bibinfo {pages} {007} (\bibinfo {year} {2019})},\ \Eprint
  {http://arxiv.org/abs/1812.05605} {arXiv:1812.05605 [hep-ph]} \BibitemShut
  {NoStop}%
\bibitem [{\citenamefont {Sabti}\ \emph {et~al.}(2020)\citenamefont {Sabti},
  \citenamefont {Alvey}, \citenamefont {Escudero}, \citenamefont {Fairbairn},\
  and\ \citenamefont {Blas}}]{Sabti:2019mhn}%
  \BibitemOpen
  \bibfield  {author} {\bibinfo {author} {\bibfnamefont {N.}~\bibnamefont
  {Sabti}}, \bibinfo {author} {\bibfnamefont {J.}~\bibnamefont {Alvey}},
  \bibinfo {author} {\bibfnamefont {M.}~\bibnamefont {Escudero}}, \bibinfo
  {author} {\bibfnamefont {M.}~\bibnamefont {Fairbairn}}, \ and\ \bibinfo
  {author} {\bibfnamefont {D.}~\bibnamefont {Blas}},\ }\href {\doibase
  10.1088/1475-7516/2020/01/004} {\bibfield  {journal} {\bibinfo  {journal}
  {JCAP}\ }\textbf {\bibinfo {volume} {01}},\ \bibinfo {pages} {004} (\bibinfo
  {year} {2020})},\ \Eprint {http://arxiv.org/abs/1910.01649} {arXiv:1910.01649
  [hep-ph]} \BibitemShut {NoStop}%
\bibitem [{\citenamefont {Hooper}\ and\ \citenamefont
  {Wang}(2004)}]{Hooper:2004qf}%
  \BibitemOpen
  \bibfield  {author} {\bibinfo {author} {\bibfnamefont {D.}~\bibnamefont
  {Hooper}}\ and\ \bibinfo {author} {\bibfnamefont {L.-T.}\ \bibnamefont
  {Wang}},\ }\href {\doibase 10.1103/PhysRevD.70.063506} {\bibfield  {journal}
  {\bibinfo  {journal} {Phys. Rev. D}\ }\textbf {\bibinfo {volume} {70}},\
  \bibinfo {pages} {063506} (\bibinfo {year} {2004})},\ \Eprint
  {http://arxiv.org/abs/hep-ph/0402220} {arXiv:hep-ph/0402220} \BibitemShut
  {NoStop}%
\bibitem [{\citenamefont {Cembranos}\ and\ \citenamefont
  {Strigari}(2008)}]{Cembranos:2008bw}%
  \BibitemOpen
  \bibfield  {author} {\bibinfo {author} {\bibfnamefont {J.~A.~R.}\
  \bibnamefont {Cembranos}}\ and\ \bibinfo {author} {\bibfnamefont {L.~E.}\
  \bibnamefont {Strigari}},\ }\href {\doibase 10.1103/PhysRevD.77.123519}
  {\bibfield  {journal} {\bibinfo  {journal} {Phys. Rev. D}\ }\textbf {\bibinfo
  {volume} {77}},\ \bibinfo {pages} {123519} (\bibinfo {year} {2008})},\
  \Eprint {http://arxiv.org/abs/0801.0630} {arXiv:0801.0630 [astro-ph]}
  \BibitemShut {NoStop}%
\bibitem [{\citenamefont {Craig}\ and\ \citenamefont
  {Raby}(2009)}]{Craig:2009zv}%
  \BibitemOpen
  \bibfield  {author} {\bibinfo {author} {\bibfnamefont {N.~J.}\ \bibnamefont
  {Craig}}\ and\ \bibinfo {author} {\bibfnamefont {S.}~\bibnamefont {Raby}},\
  }\href@noop {} {\  (\bibinfo {year} {2009})},\ \Eprint
  {http://arxiv.org/abs/0908.1842} {arXiv:0908.1842 [hep-ph]} \BibitemShut
  {NoStop}%
\bibitem [{\citenamefont {Finkbeiner}\ and\ \citenamefont
  {Weiner}(2007)}]{Finkbeiner:2007kk}%
  \BibitemOpen
  \bibfield  {author} {\bibinfo {author} {\bibfnamefont {D.~P.}\ \bibnamefont
  {Finkbeiner}}\ and\ \bibinfo {author} {\bibfnamefont {N.}~\bibnamefont
  {Weiner}},\ }\href {\doibase 10.1103/PhysRevD.76.083519} {\bibfield
  {journal} {\bibinfo  {journal} {Phys. Rev. D}\ }\textbf {\bibinfo {volume}
  {76}},\ \bibinfo {pages} {083519} (\bibinfo {year} {2007})},\ \Eprint
  {http://arxiv.org/abs/astro-ph/0702587} {arXiv:astro-ph/0702587} \BibitemShut
  {NoStop}%
\bibitem [{\citenamefont {Pospelov}\ and\ \citenamefont
  {Ritz}(2007)}]{Pospelov:2007xh}%
  \BibitemOpen
  \bibfield  {author} {\bibinfo {author} {\bibfnamefont {M.}~\bibnamefont
  {Pospelov}}\ and\ \bibinfo {author} {\bibfnamefont {A.}~\bibnamefont
  {Ritz}},\ }\href {\doibase 10.1016/j.physletb.2007.06.027} {\bibfield
  {journal} {\bibinfo  {journal} {Phys. Lett. B}\ }\textbf {\bibinfo {volume}
  {651}},\ \bibinfo {pages} {208} (\bibinfo {year} {2007})},\ \Eprint
  {http://arxiv.org/abs/hep-ph/0703128} {arXiv:hep-ph/0703128} \BibitemShut
  {NoStop}%
\bibitem [{\citenamefont {Cline}\ \emph {et~al.}(2011)\citenamefont {Cline},
  \citenamefont {Frey},\ and\ \citenamefont {Chen}}]{Cline:2010kv}%
  \BibitemOpen
  \bibfield  {author} {\bibinfo {author} {\bibfnamefont {J.~M.}\ \bibnamefont
  {Cline}}, \bibinfo {author} {\bibfnamefont {A.~R.}\ \bibnamefont {Frey}}, \
  and\ \bibinfo {author} {\bibfnamefont {F.}~\bibnamefont {Chen}},\ }\href
  {\doibase 10.1103/PhysRevD.83.083511} {\bibfield  {journal} {\bibinfo
  {journal} {Phys. Rev. D}\ }\textbf {\bibinfo {volume} {83}},\ \bibinfo
  {pages} {083511} (\bibinfo {year} {2011})},\ \Eprint
  {http://arxiv.org/abs/1008.1784} {arXiv:1008.1784 [hep-ph]} \BibitemShut
  {NoStop}%
\bibitem [{\citenamefont {Cline}\ and\ \citenamefont
  {Frey}(2012)}]{Cline:2012yx}%
  \BibitemOpen
  \bibfield  {author} {\bibinfo {author} {\bibfnamefont {J.~M.}\ \bibnamefont
  {Cline}}\ and\ \bibinfo {author} {\bibfnamefont {A.~R.}\ \bibnamefont
  {Frey}},\ }\href {\doibase 10.1002/andp.201200082} {\bibfield  {journal}
  {\bibinfo  {journal} {Annalen Phys.}\ }\textbf {\bibinfo {volume} {524}},\
  \bibinfo {pages} {579} (\bibinfo {year} {2012})},\ \Eprint
  {http://arxiv.org/abs/1204.1965} {arXiv:1204.1965 [hep-ph]} \BibitemShut
  {NoStop}%
\bibitem [{\citenamefont {Kasuya}\ and\ \citenamefont
  {Takahashi}(2005)}]{Kasuya:2005ay}%
  \BibitemOpen
  \bibfield  {author} {\bibinfo {author} {\bibfnamefont {S.}~\bibnamefont
  {Kasuya}}\ and\ \bibinfo {author} {\bibfnamefont {F.}~\bibnamefont
  {Takahashi}},\ }\href {\doibase 10.1103/PhysRevD.72.085015} {\bibfield
  {journal} {\bibinfo  {journal} {Phys. Rev. D}\ }\textbf {\bibinfo {volume}
  {72}},\ \bibinfo {pages} {085015} (\bibinfo {year} {2005})},\ \Eprint
  {http://arxiv.org/abs/astro-ph/0508391} {arXiv:astro-ph/0508391} \BibitemShut
  {NoStop}%
\bibitem [{\citenamefont {Farzan}\ and\ \citenamefont
  {Rajaee}(2017)}]{Farzan:2017hol}%
  \BibitemOpen
  \bibfield  {author} {\bibinfo {author} {\bibfnamefont {Y.}~\bibnamefont
  {Farzan}}\ and\ \bibinfo {author} {\bibfnamefont {M.}~\bibnamefont
  {Rajaee}},\ }\href {\doibase 10.1007/JHEP12(2017)083} {\bibfield  {journal}
  {\bibinfo  {journal} {JHEP}\ }\textbf {\bibinfo {volume} {12}},\ \bibinfo
  {pages} {083} (\bibinfo {year} {2017})},\ \Eprint
  {http://arxiv.org/abs/1708.01137} {arXiv:1708.01137 [hep-ph]} \BibitemShut
  {NoStop}%
\bibitem [{\citenamefont {Farzan}\ and\ \citenamefont
  {Rajaee}(2020)}]{Farzan:2020llg}%
  \BibitemOpen
  \bibfield  {author} {\bibinfo {author} {\bibfnamefont {Y.}~\bibnamefont
  {Farzan}}\ and\ \bibinfo {author} {\bibfnamefont {M.}~\bibnamefont
  {Rajaee}},\ }\href {\doibase 10.1103/PhysRevD.102.103532} {\bibfield
  {journal} {\bibinfo  {journal} {Phys. Rev. D}\ }\textbf {\bibinfo {volume}
  {102}},\ \bibinfo {pages} {103532} (\bibinfo {year} {2020})},\ \Eprint
  {http://arxiv.org/abs/2007.14421} {arXiv:2007.14421 [hep-ph]} \BibitemShut
  {NoStop}%
\bibitem [{\citenamefont {Lawson}\ and\ \citenamefont
  {Zhitnitsky}(2017)}]{Lawson:2016mpu}%
  \BibitemOpen
  \bibfield  {author} {\bibinfo {author} {\bibfnamefont {K.}~\bibnamefont
  {Lawson}}\ and\ \bibinfo {author} {\bibfnamefont {A.}~\bibnamefont
  {Zhitnitsky}},\ }\href {\doibase 10.1088/1475-7516/2017/02/049} {\bibfield
  {journal} {\bibinfo  {journal} {JCAP}\ }\textbf {\bibinfo {volume} {02}},\
  \bibinfo {pages} {049} (\bibinfo {year} {2017})},\ \Eprint
  {http://arxiv.org/abs/1611.05900} {arXiv:1611.05900 [astro-ph.CO]}
  \BibitemShut {NoStop}%
\bibitem [{\citenamefont {Davoudiasl}\ and\ \citenamefont
  {Perez}(2010)}]{Davoudiasl:2009ud}%
  \BibitemOpen
  \bibfield  {author} {\bibinfo {author} {\bibfnamefont {H.}~\bibnamefont
  {Davoudiasl}}\ and\ \bibinfo {author} {\bibfnamefont {G.}~\bibnamefont
  {Perez}},\ }\href {\doibase 10.1007/JHEP04(2010)058} {\bibfield  {journal}
  {\bibinfo  {journal} {JHEP}\ }\textbf {\bibinfo {volume} {04}},\ \bibinfo
  {pages} {058} (\bibinfo {year} {2010})},\ \Eprint
  {http://arxiv.org/abs/0912.3320} {arXiv:0912.3320 [hep-ph]} \BibitemShut
  {NoStop}%
\bibitem [{\citenamefont {Hawking}(1975)}]{Hawking:1974sw}%
  \BibitemOpen
  \bibfield  {author} {\bibinfo {author} {\bibfnamefont {S.~W.}\ \bibnamefont
  {Hawking}},\ }\href {\doibase 10.1007/BF02345020} {\bibfield  {journal}
  {\bibinfo  {journal} {Commun. Math. Phys.}\ }\textbf {\bibinfo {volume}
  {43}},\ \bibinfo {pages} {199} (\bibinfo {year} {1975})},\ \bibinfo {note}
  {[Erratum: Commun.Math.Phys. 46, 206 (1976)]}\BibitemShut {NoStop}%
\bibitem [{\citenamefont {Gibbons}\ and\ \citenamefont
  {Hawking}(1977)}]{Gibbons:1977mu}%
  \BibitemOpen
  \bibfield  {author} {\bibinfo {author} {\bibfnamefont {G.~W.}\ \bibnamefont
  {Gibbons}}\ and\ \bibinfo {author} {\bibfnamefont {S.~W.}\ \bibnamefont
  {Hawking}},\ }\href {\doibase 10.1103/PhysRevD.15.2738} {\bibfield  {journal}
  {\bibinfo  {journal} {Phys. Rev. D}\ }\textbf {\bibinfo {volume} {15}},\
  \bibinfo {pages} {2738} (\bibinfo {year} {1977})}\BibitemShut {NoStop}%
\bibitem [{\citenamefont {Frampton}\ and\ \citenamefont
  {Kephart}(2005)}]{Frampton:2005fk}%
  \BibitemOpen
  \bibfield  {author} {\bibinfo {author} {\bibfnamefont {P.~H.}\ \bibnamefont
  {Frampton}}\ and\ \bibinfo {author} {\bibfnamefont {T.~W.}\ \bibnamefont
  {Kephart}},\ }\href {\doibase 10.1142/S0217732305017688} {\bibfield
  {journal} {\bibinfo  {journal} {Mod. Phys. Lett. A}\ }\textbf {\bibinfo
  {volume} {20}},\ \bibinfo {pages} {1573} (\bibinfo {year} {2005})},\ \Eprint
  {http://arxiv.org/abs/hep-ph/0503267} {arXiv:hep-ph/0503267} \BibitemShut
  {NoStop}%
\bibitem [{\citenamefont {Bambi}\ \emph {et~al.}(2008)\citenamefont {Bambi},
  \citenamefont {Dolgov},\ and\ \citenamefont {Petrov}}]{Bambi:2008kx}%
  \BibitemOpen
  \bibfield  {author} {\bibinfo {author} {\bibfnamefont {C.}~\bibnamefont
  {Bambi}}, \bibinfo {author} {\bibfnamefont {A.~D.}\ \bibnamefont {Dolgov}}, \
  and\ \bibinfo {author} {\bibfnamefont {A.~A.}\ \bibnamefont {Petrov}},\
  }\href {\doibase 10.1016/j.physletb.2009.10.053} {\bibfield  {journal}
  {\bibinfo  {journal} {Phys. Lett. B}\ }\textbf {\bibinfo {volume} {670}},\
  \bibinfo {pages} {174} (\bibinfo {year} {2008})},\ \bibinfo {note} {[Erratum:
  Phys.Lett.B 681, 504--504 (2009)]},\ \Eprint {http://arxiv.org/abs/0801.2786}
  {arXiv:0801.2786 [astro-ph]} \BibitemShut {NoStop}%
\bibitem [{\citenamefont {Cai}\ \emph {et~al.}(2020)\citenamefont {Cai},
  \citenamefont {Ding}, \citenamefont {Yang},\ and\ \citenamefont
  {Zhou}}]{Cai:2020fnq}%
  \BibitemOpen
  \bibfield  {author} {\bibinfo {author} {\bibfnamefont {R.-G.}\ \bibnamefont
  {Cai}}, \bibinfo {author} {\bibfnamefont {Y.-C.}\ \bibnamefont {Ding}},
  \bibinfo {author} {\bibfnamefont {X.-Y.}\ \bibnamefont {Yang}}, \ and\
  \bibinfo {author} {\bibfnamefont {Y.-F.}\ \bibnamefont {Zhou}},\ }\href@noop
  {} {\  (\bibinfo {year} {2020})},\ \Eprint {http://arxiv.org/abs/2007.11804}
  {arXiv:2007.11804 [astro-ph.CO]} \BibitemShut {NoStop}%
\bibitem [{\citenamefont {Laha}(2019)}]{Laha:2019ssq}%
  \BibitemOpen
  \bibfield  {author} {\bibinfo {author} {\bibfnamefont {R.}~\bibnamefont
  {Laha}},\ }\href {\doibase 10.1103/PhysRevLett.123.251101} {\bibfield
  {journal} {\bibinfo  {journal} {Phys. Rev. Lett.}\ }\textbf {\bibinfo
  {volume} {123}},\ \bibinfo {pages} {251101} (\bibinfo {year} {2019})},\
  \Eprint {http://arxiv.org/abs/1906.09994} {arXiv:1906.09994 [astro-ph.HE]}
  \BibitemShut {NoStop}%
\bibitem [{\citenamefont {DeRocco}\ and\ \citenamefont
  {Graham}(2019)}]{DeRocco:2019fjq}%
  \BibitemOpen
  \bibfield  {author} {\bibinfo {author} {\bibfnamefont {W.}~\bibnamefont
  {DeRocco}}\ and\ \bibinfo {author} {\bibfnamefont {P.~W.}\ \bibnamefont
  {Graham}},\ }\href {\doibase 10.1103/PhysRevLett.123.251102} {\bibfield
  {journal} {\bibinfo  {journal} {Phys. Rev. Lett.}\ }\textbf {\bibinfo
  {volume} {123}},\ \bibinfo {pages} {251102} (\bibinfo {year} {2019})},\
  \Eprint {http://arxiv.org/abs/1906.07740} {arXiv:1906.07740 [astro-ph.CO]}
  \BibitemShut {NoStop}%
\bibitem [{\citenamefont {Laha}\ \emph {et~al.}(2020)\citenamefont {Laha},
  \citenamefont {Mu\~noz},\ and\ \citenamefont {Slatyer}}]{Laha:2020ivk}%
  \BibitemOpen
  \bibfield  {author} {\bibinfo {author} {\bibfnamefont {R.}~\bibnamefont
  {Laha}}, \bibinfo {author} {\bibfnamefont {J.~B.}\ \bibnamefont {Mu\~noz}}, \
  and\ \bibinfo {author} {\bibfnamefont {T.~R.}\ \bibnamefont {Slatyer}},\
  }\href {\doibase 10.1103/PhysRevD.101.123514} {\bibfield  {journal} {\bibinfo
   {journal} {Phys. Rev. D}\ }\textbf {\bibinfo {volume} {101}},\ \bibinfo
  {pages} {123514} (\bibinfo {year} {2020})},\ \Eprint
  {http://arxiv.org/abs/2004.00627} {arXiv:2004.00627 [astro-ph.CO]}
  \BibitemShut {NoStop}%
\bibitem [{\citenamefont {Coogan}\ \emph {et~al.}(2020)\citenamefont {Coogan},
  \citenamefont {Morrison},\ and\ \citenamefont {Profumo}}]{Coogan:2020tuf}%
  \BibitemOpen
  \bibfield  {author} {\bibinfo {author} {\bibfnamefont {A.}~\bibnamefont
  {Coogan}}, \bibinfo {author} {\bibfnamefont {L.}~\bibnamefont {Morrison}}, \
  and\ \bibinfo {author} {\bibfnamefont {S.}~\bibnamefont {Profumo}},\
  }\href@noop {} {\  (\bibinfo {year} {2020})},\ \Eprint
  {http://arxiv.org/abs/2010.04797} {arXiv:2010.04797 [astro-ph.CO]}
  \BibitemShut {NoStop}%
\bibitem [{\citenamefont {Boudaud}\ and\ \citenamefont
  {Cirelli}(2019)}]{Boudaud:2018hqb}%
  \BibitemOpen
  \bibfield  {author} {\bibinfo {author} {\bibfnamefont {M.}~\bibnamefont
  {Boudaud}}\ and\ \bibinfo {author} {\bibfnamefont {M.}~\bibnamefont
  {Cirelli}},\ }\href {\doibase 10.1103/PhysRevLett.122.041104} {\bibfield
  {journal} {\bibinfo  {journal} {Phys. Rev. Lett.}\ }\textbf {\bibinfo
  {volume} {122}},\ \bibinfo {pages} {041104} (\bibinfo {year} {2019})},\
  \Eprint {http://arxiv.org/abs/1807.03075} {arXiv:1807.03075 [astro-ph.HE]}
  \BibitemShut {NoStop}%
\bibitem [{\citenamefont {Carr}\ \emph {et~al.}(2010)\citenamefont {Carr},
  \citenamefont {Kohri}, \citenamefont {Sendouda},\ and\ \citenamefont
  {Yokoyama}}]{Carr:2009jm}%
  \BibitemOpen
  \bibfield  {author} {\bibinfo {author} {\bibfnamefont {B.~J.}\ \bibnamefont
  {Carr}}, \bibinfo {author} {\bibfnamefont {K.}~\bibnamefont {Kohri}},
  \bibinfo {author} {\bibfnamefont {Y.}~\bibnamefont {Sendouda}}, \ and\
  \bibinfo {author} {\bibfnamefont {J.}~\bibnamefont {Yokoyama}},\ }\href
  {\doibase 10.1103/PhysRevD.81.104019} {\bibfield  {journal} {\bibinfo
  {journal} {Phys. Rev. D}\ }\textbf {\bibinfo {volume} {81}},\ \bibinfo
  {pages} {104019} (\bibinfo {year} {2010})},\ \Eprint
  {http://arxiv.org/abs/0912.5297} {arXiv:0912.5297 [astro-ph.CO]} \BibitemShut
  {NoStop}%
\bibitem [{\citenamefont {Carr}\ \emph
  {et~al.}(2020{\natexlab{a}})\citenamefont {Carr}, \citenamefont {Kohri},
  \citenamefont {Sendouda},\ and\ \citenamefont {Yokoyama}}]{Carr:2020gox}%
  \BibitemOpen
  \bibfield  {author} {\bibinfo {author} {\bibfnamefont {B.}~\bibnamefont
  {Carr}}, \bibinfo {author} {\bibfnamefont {K.}~\bibnamefont {Kohri}},
  \bibinfo {author} {\bibfnamefont {Y.}~\bibnamefont {Sendouda}}, \ and\
  \bibinfo {author} {\bibfnamefont {J.}~\bibnamefont {Yokoyama}},\ }\href@noop
  {} {\  (\bibinfo {year} {2020}{\natexlab{a}})},\ \Eprint
  {http://arxiv.org/abs/2002.12778} {arXiv:2002.12778 [astro-ph.CO]}
  \BibitemShut {NoStop}%
\bibitem [{\citenamefont {Dasgupta}\ \emph {et~al.}(2020)\citenamefont
  {Dasgupta}, \citenamefont {Laha},\ and\ \citenamefont
  {Ray}}]{Dasgupta:2019cae}%
  \BibitemOpen
  \bibfield  {author} {\bibinfo {author} {\bibfnamefont {B.}~\bibnamefont
  {Dasgupta}}, \bibinfo {author} {\bibfnamefont {R.}~\bibnamefont {Laha}}, \
  and\ \bibinfo {author} {\bibfnamefont {A.}~\bibnamefont {Ray}},\ }\href
  {\doibase 10.1103/PhysRevLett.125.101101} {\bibfield  {journal} {\bibinfo
  {journal} {Phys. Rev. Lett.}\ }\textbf {\bibinfo {volume} {125}},\ \bibinfo
  {pages} {101101} (\bibinfo {year} {2020})},\ \Eprint
  {http://arxiv.org/abs/1912.01014} {arXiv:1912.01014 [hep-ph]} \BibitemShut
  {NoStop}%
\bibitem [{\citenamefont {Bouchet}\ \emph {et~al.}(2010)\citenamefont
  {Bouchet}, \citenamefont {Roques},\ and\ \citenamefont
  {Jourdain}}]{Bouchet:2010dj}%
  \BibitemOpen
  \bibfield  {author} {\bibinfo {author} {\bibfnamefont {L.}~\bibnamefont
  {Bouchet}}, \bibinfo {author} {\bibfnamefont {J.-P.}\ \bibnamefont {Roques}},
  \ and\ \bibinfo {author} {\bibfnamefont {E.}~\bibnamefont {Jourdain}},\
  }\href {\doibase 10.1088/0004-637X/720/2/1772} {\bibfield  {journal}
  {\bibinfo  {journal} {Astrophys. J.}\ }\textbf {\bibinfo {volume} {720}},\
  \bibinfo {pages} {1772} (\bibinfo {year} {2010})},\ \Eprint
  {http://arxiv.org/abs/1007.4753} {arXiv:1007.4753 [astro-ph.HE]} \BibitemShut
  {NoStop}%
\bibitem [{\citenamefont {Robin}\ \emph {et~al.}(2003)\citenamefont {Robin},
  \citenamefont {Reyle}, \citenamefont {Derriere},\ and\ \citenamefont
  {Picaud}}]{Robin:2004qd}%
  \BibitemOpen
  \bibfield  {author} {\bibinfo {author} {\bibfnamefont {A.~C.}\ \bibnamefont
  {Robin}}, \bibinfo {author} {\bibfnamefont {C.}~\bibnamefont {Reyle}},
  \bibinfo {author} {\bibfnamefont {S.}~\bibnamefont {Derriere}}, \ and\
  \bibinfo {author} {\bibfnamefont {S.}~\bibnamefont {Picaud}},\ }\href
  {\doibase 10.1051/0004-6361:20040968} {\bibfield  {journal} {\bibinfo
  {journal} {Astron. Astrophys.}\ }\textbf {\bibinfo {volume} {409}},\ \bibinfo
  {pages} {523} (\bibinfo {year} {2003})},\ \Eprint
  {http://arxiv.org/abs/astro-ph/0401052} {arXiv:astro-ph/0401052} \BibitemShut
  {NoStop}%
\bibitem [{\citenamefont {MacGibbon}\ and\ \citenamefont
  {Webber}(1990)}]{MacGibbon:1990zk}%
  \BibitemOpen
  \bibfield  {author} {\bibinfo {author} {\bibfnamefont {J.~H.}\ \bibnamefont
  {MacGibbon}}\ and\ \bibinfo {author} {\bibfnamefont {B.~R.}\ \bibnamefont
  {Webber}},\ }\href {\doibase 10.1103/PhysRevD.41.3052} {\bibfield  {journal}
  {\bibinfo  {journal} {Phys. Rev. D}\ }\textbf {\bibinfo {volume} {41}},\
  \bibinfo {pages} {3052} (\bibinfo {year} {1990})}\BibitemShut {NoStop}%
\bibitem [{\citenamefont {MacGibbon}(1991)}]{MacGibbon:1991tj}%
  \BibitemOpen
  \bibfield  {author} {\bibinfo {author} {\bibfnamefont {J.~H.}\ \bibnamefont
  {MacGibbon}},\ }\href {\doibase 10.1103/PhysRevD.44.376} {\bibfield
  {journal} {\bibinfo  {journal} {Phys. Rev. D}\ }\textbf {\bibinfo {volume}
  {44}},\ \bibinfo {pages} {376} (\bibinfo {year} {1991})}\BibitemShut
  {NoStop}%
\bibitem [{\citenamefont {Page}(1976)}]{Page:1976df}%
  \BibitemOpen
  \bibfield  {author} {\bibinfo {author} {\bibfnamefont {D.~N.}\ \bibnamefont
  {Page}},\ }\href {\doibase 10.1103/PhysRevD.13.198} {\bibfield  {journal}
  {\bibinfo  {journal} {Phys. Rev. D}\ }\textbf {\bibinfo {volume} {13}},\
  \bibinfo {pages} {198} (\bibinfo {year} {1976})}\BibitemShut {NoStop}%
\bibitem [{\citenamefont {Navarro}\ \emph {et~al.}(1996)\citenamefont
  {Navarro}, \citenamefont {Frenk},\ and\ \citenamefont
  {White}}]{Navarro:1995iw}%
  \BibitemOpen
  \bibfield  {author} {\bibinfo {author} {\bibfnamefont {J.~F.}\ \bibnamefont
  {Navarro}}, \bibinfo {author} {\bibfnamefont {C.~S.}\ \bibnamefont {Frenk}},
  \ and\ \bibinfo {author} {\bibfnamefont {S.~D.~M.}\ \bibnamefont {White}},\
  }\href {\doibase 10.1086/177173} {\bibfield  {journal} {\bibinfo  {journal}
  {Astrophys. J.}\ }\textbf {\bibinfo {volume} {462}},\ \bibinfo {pages} {563}
  (\bibinfo {year} {1996})},\ \Eprint {http://arxiv.org/abs/astro-ph/9508025}
  {arXiv:astro-ph/9508025} \BibitemShut {NoStop}%
\bibitem [{\citenamefont {Navarro}\ \emph {et~al.}(1997)\citenamefont
  {Navarro}, \citenamefont {Frenk},\ and\ \citenamefont
  {White}}]{Navarro:1996gj}%
  \BibitemOpen
  \bibfield  {author} {\bibinfo {author} {\bibfnamefont {J.~F.}\ \bibnamefont
  {Navarro}}, \bibinfo {author} {\bibfnamefont {C.~S.}\ \bibnamefont {Frenk}},
  \ and\ \bibinfo {author} {\bibfnamefont {S.~D.~M.}\ \bibnamefont {White}},\
  }\href {\doibase 10.1086/304888} {\bibfield  {journal} {\bibinfo  {journal}
  {Astrophys. J.}\ }\textbf {\bibinfo {volume} {490}},\ \bibinfo {pages} {493}
  (\bibinfo {year} {1997})},\ \Eprint {http://arxiv.org/abs/astro-ph/9611107}
  {arXiv:astro-ph/9611107} \BibitemShut {NoStop}%
\bibitem [{\citenamefont {Vincent}\ \emph {et~al.}(2012)\citenamefont
  {Vincent}, \citenamefont {Martin},\ and\ \citenamefont
  {Cline}}]{Vincent:2012an}%
  \BibitemOpen
  \bibfield  {author} {\bibinfo {author} {\bibfnamefont {A.~C.}\ \bibnamefont
  {Vincent}}, \bibinfo {author} {\bibfnamefont {P.}~\bibnamefont {Martin}}, \
  and\ \bibinfo {author} {\bibfnamefont {J.~M.}\ \bibnamefont {Cline}},\ }\href
  {\doibase 10.1088/1475-7516/2012/04/022} {\bibfield  {journal} {\bibinfo
  {journal} {JCAP}\ }\textbf {\bibinfo {volume} {04}},\ \bibinfo {pages} {022}
  (\bibinfo {year} {2012})},\ \Eprint {http://arxiv.org/abs/1201.0997}
  {arXiv:1201.0997 [hep-ph]} \BibitemShut {NoStop}%
\bibitem [{\citenamefont {Ascasibar}\ \emph {et~al.}(2006)\citenamefont
  {Ascasibar}, \citenamefont {Jean}, \citenamefont {Boehm},\ and\ \citenamefont
  {Knoedlseder}}]{Ascasibar:2005rw}%
  \BibitemOpen
  \bibfield  {author} {\bibinfo {author} {\bibfnamefont {Y.}~\bibnamefont
  {Ascasibar}}, \bibinfo {author} {\bibfnamefont {P.}~\bibnamefont {Jean}},
  \bibinfo {author} {\bibfnamefont {C.}~\bibnamefont {Boehm}}, \ and\ \bibinfo
  {author} {\bibfnamefont {J.}~\bibnamefont {Knoedlseder}},\ }\href {\doibase
  10.1111/j.1365-2966.2006.10226.x} {\bibfield  {journal} {\bibinfo  {journal}
  {Mon. Not. Roy. Astron. Soc.}\ }\textbf {\bibinfo {volume} {368}},\ \bibinfo
  {pages} {1695} (\bibinfo {year} {2006})},\ \Eprint
  {http://arxiv.org/abs/astro-ph/0507142} {arXiv:astro-ph/0507142} \BibitemShut
  {NoStop}%
\bibitem [{\citenamefont {Gnedin}\ \emph
  {et~al.}(2011{\natexlab{a}})\citenamefont {Gnedin}, \citenamefont {Ceverino},
  \citenamefont {Gnedin}, \citenamefont {Klypin}, \citenamefont {Kravtsov},
  \citenamefont {Levine}, \citenamefont {Nagai},\ and\ \citenamefont
  {Yepes}}]{gnedin2011halo}%
  \BibitemOpen
  \bibfield  {author} {\bibinfo {author} {\bibfnamefont {O.~Y.}\ \bibnamefont
  {Gnedin}}, \bibinfo {author} {\bibfnamefont {D.}~\bibnamefont {Ceverino}},
  \bibinfo {author} {\bibfnamefont {N.~Y.}\ \bibnamefont {Gnedin}}, \bibinfo
  {author} {\bibfnamefont {A.~A.}\ \bibnamefont {Klypin}}, \bibinfo {author}
  {\bibfnamefont {A.~V.}\ \bibnamefont {Kravtsov}}, \bibinfo {author}
  {\bibfnamefont {R.}~\bibnamefont {Levine}}, \bibinfo {author} {\bibfnamefont
  {D.}~\bibnamefont {Nagai}}, \ and\ \bibinfo {author} {\bibfnamefont
  {G.}~\bibnamefont {Yepes}},\ }\href@noop {} {\enquote {\bibinfo {title} {Halo
  contraction effect in hydrodynamic simulations of galaxy formation},}\ }
  (\bibinfo {year} {2011}{\natexlab{a}}),\ \Eprint
  {http://arxiv.org/abs/1108.5736} {arXiv:1108.5736 [astro-ph.CO]} \BibitemShut
  {NoStop}%
\bibitem [{\citenamefont {Gnedin}\ \emph {et~al.}(2004)\citenamefont {Gnedin},
  \citenamefont {Kravtsov}, \citenamefont {Klypin},\ and\ \citenamefont
  {Nagai}}]{Gnedin:2004cx}%
  \BibitemOpen
  \bibfield  {author} {\bibinfo {author} {\bibfnamefont {O.~Y.}\ \bibnamefont
  {Gnedin}}, \bibinfo {author} {\bibfnamefont {A.~V.}\ \bibnamefont
  {Kravtsov}}, \bibinfo {author} {\bibfnamefont {A.~A.}\ \bibnamefont
  {Klypin}}, \ and\ \bibinfo {author} {\bibfnamefont {D.}~\bibnamefont
  {Nagai}},\ }\href {\doibase 10.1086/424914} {\bibfield  {journal} {\bibinfo
  {journal} {Astrophys. J.}\ }\textbf {\bibinfo {volume} {616}},\ \bibinfo
  {pages} {16} (\bibinfo {year} {2004})},\ \Eprint
  {http://arxiv.org/abs/astro-ph/0406247} {arXiv:astro-ph/0406247} \BibitemShut
  {NoStop}%
\bibitem [{\citenamefont {Governato}\ \emph {et~al.}(2012)\citenamefont
  {Governato}, \citenamefont {Zolotov}, \citenamefont {Pontzen}, \citenamefont
  {Christensen}, \citenamefont {Oh}, \citenamefont {Brooks}, \citenamefont
  {Quinn}, \citenamefont {Shen},\ and\ \citenamefont
  {Wadsley}}]{Governato_2012}%
  \BibitemOpen
  \bibfield  {author} {\bibinfo {author} {\bibfnamefont {F.}~\bibnamefont
  {Governato}}, \bibinfo {author} {\bibfnamefont {A.}~\bibnamefont {Zolotov}},
  \bibinfo {author} {\bibfnamefont {A.}~\bibnamefont {Pontzen}}, \bibinfo
  {author} {\bibfnamefont {C.}~\bibnamefont {Christensen}}, \bibinfo {author}
  {\bibfnamefont {S.~H.}\ \bibnamefont {Oh}}, \bibinfo {author} {\bibfnamefont
  {A.~M.}\ \bibnamefont {Brooks}}, \bibinfo {author} {\bibfnamefont
  {T.}~\bibnamefont {Quinn}}, \bibinfo {author} {\bibfnamefont
  {S.}~\bibnamefont {Shen}}, \ and\ \bibinfo {author} {\bibfnamefont
  {J.}~\bibnamefont {Wadsley}},\ }\href {\doibase
  10.1111/j.1365-2966.2012.20696.x} {\bibfield  {journal} {\bibinfo  {journal}
  {Monthly Notices of the Royal Astronomical Society}\ }\textbf {\bibinfo
  {volume} {422}},\ \bibinfo {pages} {1231?1240} (\bibinfo {year}
  {2012})}\BibitemShut {NoStop}%
\bibitem [{\citenamefont {Kuhlen}\ \emph {et~al.}(2013)\citenamefont {Kuhlen},
  \citenamefont {Guedes}, \citenamefont {Pillepich}, \citenamefont {Madau},\
  and\ \citenamefont {Mayer}}]{Kuhlen:2012qw}%
  \BibitemOpen
  \bibfield  {author} {\bibinfo {author} {\bibfnamefont {M.}~\bibnamefont
  {Kuhlen}}, \bibinfo {author} {\bibfnamefont {J.}~\bibnamefont {Guedes}},
  \bibinfo {author} {\bibfnamefont {A.}~\bibnamefont {Pillepich}}, \bibinfo
  {author} {\bibfnamefont {P.}~\bibnamefont {Madau}}, \ and\ \bibinfo {author}
  {\bibfnamefont {L.}~\bibnamefont {Mayer}},\ }\href {\doibase
  10.1088/0004-637X/765/1/10} {\bibfield  {journal} {\bibinfo  {journal}
  {Astrophys. J.}\ }\textbf {\bibinfo {volume} {765}},\ \bibinfo {pages} {10}
  (\bibinfo {year} {2013})},\ \Eprint {http://arxiv.org/abs/1208.4844}
  {arXiv:1208.4844 [astro-ph.GA]} \BibitemShut {NoStop}%
\bibitem [{\citenamefont {Weinberg}\ and\ \citenamefont
  {Katz}(2002)}]{Weinberg:2001gm}%
  \BibitemOpen
  \bibfield  {author} {\bibinfo {author} {\bibfnamefont {M.~D.}\ \bibnamefont
  {Weinberg}}\ and\ \bibinfo {author} {\bibfnamefont {N.}~\bibnamefont
  {Katz}},\ }\href {\doibase 10.1086/343847} {\bibfield  {journal} {\bibinfo
  {journal} {Astrophys. J.}\ }\textbf {\bibinfo {volume} {580}},\ \bibinfo
  {pages} {627} (\bibinfo {year} {2002})},\ \Eprint
  {http://arxiv.org/abs/astro-ph/0110632} {arXiv:astro-ph/0110632} \BibitemShut
  {NoStop}%
\bibitem [{\citenamefont {Weinberg}\ and\ \citenamefont
  {Katz}(2007)}]{Weinberg:2006ps}%
  \BibitemOpen
  \bibfield  {author} {\bibinfo {author} {\bibfnamefont {M.~D.}\ \bibnamefont
  {Weinberg}}\ and\ \bibinfo {author} {\bibfnamefont {N.}~\bibnamefont
  {Katz}},\ }\href {\doibase 10.1111/j.1365-2966.2006.11307.x} {\bibfield
  {journal} {\bibinfo  {journal} {Mon. Not. Roy. Astron. Soc.}\ }\textbf
  {\bibinfo {volume} {375}},\ \bibinfo {pages} {460} (\bibinfo {year}
  {2007})},\ \Eprint {http://arxiv.org/abs/astro-ph/0601138}
  {arXiv:astro-ph/0601138} \BibitemShut {NoStop}%
\bibitem [{\citenamefont {Sellwood}(2003)}]{Sellwood:2002vb}%
  \BibitemOpen
  \bibfield  {author} {\bibinfo {author} {\bibfnamefont {J.~A.}\ \bibnamefont
  {Sellwood}},\ }\href {\doibase 10.1086/368285} {\bibfield  {journal}
  {\bibinfo  {journal} {Astrophys. J.}\ }\textbf {\bibinfo {volume} {587}},\
  \bibinfo {pages} {638} (\bibinfo {year} {2003})},\ \Eprint
  {http://arxiv.org/abs/astro-ph/0210079} {arXiv:astro-ph/0210079} \BibitemShut
  {NoStop}%
\bibitem [{\citenamefont {Valenzuela}\ and\ \citenamefont
  {Klypin}(2003)}]{Valenzuela:2002np}%
  \BibitemOpen
  \bibfield  {author} {\bibinfo {author} {\bibfnamefont {O.}~\bibnamefont
  {Valenzuela}}\ and\ \bibinfo {author} {\bibfnamefont {A.}~\bibnamefont
  {Klypin}},\ }\href {\doibase 10.1046/j.1365-8711.2003.06947.x} {\bibfield
  {journal} {\bibinfo  {journal} {Mon. Not. Roy. Astron. Soc.}\ }\textbf
  {\bibinfo {volume} {345}},\ \bibinfo {pages} {406} (\bibinfo {year}
  {2003})},\ \Eprint {http://arxiv.org/abs/astro-ph/0204028}
  {arXiv:astro-ph/0204028} \BibitemShut {NoStop}%
\bibitem [{\citenamefont {Colin}\ \emph {et~al.}(2006)\citenamefont {Colin},
  \citenamefont {Valenzuela},\ and\ \citenamefont {Klypin}}]{Colin:2005rr}%
  \BibitemOpen
  \bibfield  {author} {\bibinfo {author} {\bibfnamefont {P.}~\bibnamefont
  {Colin}}, \bibinfo {author} {\bibfnamefont {O.}~\bibnamefont {Valenzuela}}, \
  and\ \bibinfo {author} {\bibfnamefont {A.}~\bibnamefont {Klypin}},\ }\href
  {\doibase 10.1086/503791} {\bibfield  {journal} {\bibinfo  {journal}
  {Astrophys. J.}\ }\textbf {\bibinfo {volume} {644}},\ \bibinfo {pages} {687}
  (\bibinfo {year} {2006})},\ \Eprint {http://arxiv.org/abs/astro-ph/0506627}
  {arXiv:astro-ph/0506627} \BibitemShut {NoStop}%
\bibitem [{\citenamefont {Scannapieco}\ \emph {et~al.}(2012)\citenamefont
  {Scannapieco}, \citenamefont {Wadepuhl}, \citenamefont {Parry}, \citenamefont
  {Navarro}, \citenamefont {Jenkins}, \citenamefont {Springel}, \citenamefont
  {Teyssier}, \citenamefont {Carlson}, \citenamefont {Couchman}, \citenamefont
  {Crain},\ and\ \citenamefont {et~al.}}]{Scannapieco_2012}%
  \BibitemOpen
  \bibfield  {author} {\bibinfo {author} {\bibfnamefont {C.}~\bibnamefont
  {Scannapieco}}, \bibinfo {author} {\bibfnamefont {M.}~\bibnamefont
  {Wadepuhl}}, \bibinfo {author} {\bibfnamefont {O.~H.}\ \bibnamefont {Parry}},
  \bibinfo {author} {\bibfnamefont {J.~F.}\ \bibnamefont {Navarro}}, \bibinfo
  {author} {\bibfnamefont {A.}~\bibnamefont {Jenkins}}, \bibinfo {author}
  {\bibfnamefont {V.}~\bibnamefont {Springel}}, \bibinfo {author}
  {\bibfnamefont {R.}~\bibnamefont {Teyssier}}, \bibinfo {author}
  {\bibfnamefont {E.}~\bibnamefont {Carlson}}, \bibinfo {author} {\bibfnamefont
  {H.~M.~P.}\ \bibnamefont {Couchman}}, \bibinfo {author} {\bibfnamefont
  {R.~A.}\ \bibnamefont {Crain}}, \ and\ \bibinfo {author} {\bibnamefont
  {et~al.}},\ }\href {\doibase 10.1111/j.1365-2966.2012.20993.x} {\bibfield
  {journal} {\bibinfo  {journal} {Monthly Notices of the Royal Astronomical
  Society}\ }\textbf {\bibinfo {volume} {423}},\ \bibinfo {pages} {1726?1749}
  (\bibinfo {year} {2012})}\BibitemShut {NoStop}%
\bibitem [{\citenamefont {Calore}\ \emph {et~al.}(2015)\citenamefont {Calore},
  \citenamefont {Bozorgnia}, \citenamefont {Lovell}, \citenamefont {Bertone},
  \citenamefont {Schaller}, \citenamefont {Frenk}, \citenamefont {Crain},
  \citenamefont {Schaye}, \citenamefont {Theuns},\ and\ \citenamefont
  {Trayford}}]{Calore:2015oya}%
  \BibitemOpen
  \bibfield  {author} {\bibinfo {author} {\bibfnamefont {F.}~\bibnamefont
  {Calore}}, \bibinfo {author} {\bibfnamefont {N.}~\bibnamefont {Bozorgnia}},
  \bibinfo {author} {\bibfnamefont {M.}~\bibnamefont {Lovell}}, \bibinfo
  {author} {\bibfnamefont {G.}~\bibnamefont {Bertone}}, \bibinfo {author}
  {\bibfnamefont {M.}~\bibnamefont {Schaller}}, \bibinfo {author}
  {\bibfnamefont {C.~S.}\ \bibnamefont {Frenk}}, \bibinfo {author}
  {\bibfnamefont {R.~A.}\ \bibnamefont {Crain}}, \bibinfo {author}
  {\bibfnamefont {J.}~\bibnamefont {Schaye}}, \bibinfo {author} {\bibfnamefont
  {T.}~\bibnamefont {Theuns}}, \ and\ \bibinfo {author} {\bibfnamefont {J.~W.}\
  \bibnamefont {Trayford}},\ }\href {\doibase 10.1088/1475-7516/2015/12/053}
  {\bibfield  {journal} {\bibinfo  {journal} {JCAP}\ }\textbf {\bibinfo
  {volume} {12}},\ \bibinfo {pages} {053} (\bibinfo {year} {2015})},\ \Eprint
  {http://arxiv.org/abs/1509.02164} {arXiv:1509.02164 [astro-ph.GA]}
  \BibitemShut {NoStop}%
\bibitem [{\citenamefont {Schaller}\ \emph {et~al.}(2015)\citenamefont
  {Schaller}, \citenamefont {Frenk}, \citenamefont {Bower}, \citenamefont
  {Theuns}, \citenamefont {Jenkins}, \citenamefont {Schaye}, \citenamefont
  {Crain}, \citenamefont {Furlong}, \citenamefont {Vecchia},\ and\
  \citenamefont {McCarthy}}]{Schaller:2014uwa}%
  \BibitemOpen
  \bibfield  {author} {\bibinfo {author} {\bibfnamefont {M.}~\bibnamefont
  {Schaller}}, \bibinfo {author} {\bibfnamefont {C.~S.}\ \bibnamefont {Frenk}},
  \bibinfo {author} {\bibfnamefont {R.~G.}\ \bibnamefont {Bower}}, \bibinfo
  {author} {\bibfnamefont {T.}~\bibnamefont {Theuns}}, \bibinfo {author}
  {\bibfnamefont {A.}~\bibnamefont {Jenkins}}, \bibinfo {author} {\bibfnamefont
  {J.}~\bibnamefont {Schaye}}, \bibinfo {author} {\bibfnamefont {R.~A.}\
  \bibnamefont {Crain}}, \bibinfo {author} {\bibfnamefont {M.}~\bibnamefont
  {Furlong}}, \bibinfo {author} {\bibfnamefont {C.~D.}\ \bibnamefont
  {Vecchia}}, \ and\ \bibinfo {author} {\bibfnamefont {I.~G.}\ \bibnamefont
  {McCarthy}},\ }\href {\doibase 10.1093/mnras/stv1067} {\bibfield  {journal}
  {\bibinfo  {journal} {Mon. Not. Roy. Astron. Soc.}\ }\textbf {\bibinfo
  {volume} {451}},\ \bibinfo {pages} {1247} (\bibinfo {year} {2015})},\ \Eprint
  {http://arxiv.org/abs/1409.8617} {arXiv:1409.8617 [astro-ph.CO]} \BibitemShut
  {NoStop}%
\bibitem [{\citenamefont {Di~Cintio}\ \emph
  {et~al.}(2014{\natexlab{a}})\citenamefont {Di~Cintio}, \citenamefont {Brook},
  \citenamefont {Dutton}, \citenamefont {Macci\`o}, \citenamefont {Stinson},\
  and\ \citenamefont {Knebe}}]{DiCintio:2014xia}%
  \BibitemOpen
  \bibfield  {author} {\bibinfo {author} {\bibfnamefont {A.}~\bibnamefont
  {Di~Cintio}}, \bibinfo {author} {\bibfnamefont {C.~B.}\ \bibnamefont
  {Brook}}, \bibinfo {author} {\bibfnamefont {A.~A.}\ \bibnamefont {Dutton}},
  \bibinfo {author} {\bibfnamefont {A.~V.}\ \bibnamefont {Macci\`o}}, \bibinfo
  {author} {\bibfnamefont {G.~S.}\ \bibnamefont {Stinson}}, \ and\ \bibinfo
  {author} {\bibfnamefont {A.}~\bibnamefont {Knebe}},\ }\href {\doibase
  10.1093/mnras/stu729} {\bibfield  {journal} {\bibinfo  {journal} {Mon. Not.
  Roy. Astron. Soc.}\ }\textbf {\bibinfo {volume} {441}},\ \bibinfo {pages}
  {2986} (\bibinfo {year} {2014}{\natexlab{a}})},\ \Eprint
  {http://arxiv.org/abs/1404.5959} {arXiv:1404.5959 [astro-ph.CO]} \BibitemShut
  {NoStop}%
\bibitem [{\citenamefont {Di~Cintio}\ \emph
  {et~al.}(2014{\natexlab{b}})\citenamefont {Di~Cintio}, \citenamefont {Brook},
  \citenamefont {Macci\`o}, \citenamefont {Stinson}, \citenamefont {Knebe},
  \citenamefont {Dutton},\ and\ \citenamefont {Wadsley}}]{DiCintio:2013qxa}%
  \BibitemOpen
  \bibfield  {author} {\bibinfo {author} {\bibfnamefont {A.}~\bibnamefont
  {Di~Cintio}}, \bibinfo {author} {\bibfnamefont {C.~B.}\ \bibnamefont
  {Brook}}, \bibinfo {author} {\bibfnamefont {A.~V.}\ \bibnamefont {Macci\`o}},
  \bibinfo {author} {\bibfnamefont {G.~S.}\ \bibnamefont {Stinson}}, \bibinfo
  {author} {\bibfnamefont {A.}~\bibnamefont {Knebe}}, \bibinfo {author}
  {\bibfnamefont {A.~A.}\ \bibnamefont {Dutton}}, \ and\ \bibinfo {author}
  {\bibfnamefont {J.}~\bibnamefont {Wadsley}},\ }\href {\doibase
  10.1093/mnras/stt1891} {\bibfield  {journal} {\bibinfo  {journal} {Mon. Not.
  Roy. Astron. Soc.}\ }\textbf {\bibinfo {volume} {437}},\ \bibinfo {pages}
  {415} (\bibinfo {year} {2014}{\natexlab{b}})},\ \Eprint
  {http://arxiv.org/abs/1306.0898} {arXiv:1306.0898 [astro-ph.CO]} \BibitemShut
  {NoStop}%
\bibitem [{\citenamefont {Schaller}\ \emph {et~al.}(2016)\citenamefont
  {Schaller} \emph {et~al.}}]{Schaller:2015mua}%
  \BibitemOpen
  \bibfield  {author} {\bibinfo {author} {\bibfnamefont {M.}~\bibnamefont
  {Schaller}} \emph {et~al.},\ }\href {\doibase 10.1093/mnras/stv2667}
  {\bibfield  {journal} {\bibinfo  {journal} {Mon. Not. Roy. Astron. Soc.}\
  }\textbf {\bibinfo {volume} {455}},\ \bibinfo {pages} {4442} (\bibinfo {year}
  {2016})},\ \Eprint {http://arxiv.org/abs/1509.02166} {arXiv:1509.02166
  [astro-ph.CO]} \BibitemShut {NoStop}%
\bibitem [{\citenamefont {Bernal}\ \emph {et~al.}(2016)\citenamefont {Bernal},
  \citenamefont {Necib},\ and\ \citenamefont {Slatyer}}]{Bernal:2016guq}%
  \BibitemOpen
  \bibfield  {author} {\bibinfo {author} {\bibfnamefont {N.}~\bibnamefont
  {Bernal}}, \bibinfo {author} {\bibfnamefont {L.}~\bibnamefont {Necib}}, \
  and\ \bibinfo {author} {\bibfnamefont {T.~R.}\ \bibnamefont {Slatyer}},\
  }\href {\doibase 10.1088/1475-7516/2016/12/030} {\bibfield  {journal}
  {\bibinfo  {journal} {JCAP}\ }\textbf {\bibinfo {volume} {12}},\ \bibinfo
  {pages} {030} (\bibinfo {year} {2016})},\ \Eprint
  {http://arxiv.org/abs/1606.00433} {arXiv:1606.00433 [astro-ph.CO]}
  \BibitemShut {NoStop}%
\bibitem [{\citenamefont {Gnedin}\ \emph
  {et~al.}(2011{\natexlab{b}})\citenamefont {Gnedin}, \citenamefont {Ceverino},
  \citenamefont {Gnedin}, \citenamefont {Klypin}, \citenamefont {Kravtsov},
  \citenamefont {Levine}, \citenamefont {Nagai},\ and\ \citenamefont
  {Yepes}}]{Gnedin:2011uj}%
  \BibitemOpen
  \bibfield  {author} {\bibinfo {author} {\bibfnamefont {O.~Y.}\ \bibnamefont
  {Gnedin}}, \bibinfo {author} {\bibfnamefont {D.}~\bibnamefont {Ceverino}},
  \bibinfo {author} {\bibfnamefont {N.~Y.}\ \bibnamefont {Gnedin}}, \bibinfo
  {author} {\bibfnamefont {A.~A.}\ \bibnamefont {Klypin}}, \bibinfo {author}
  {\bibfnamefont {A.~V.}\ \bibnamefont {Kravtsov}}, \bibinfo {author}
  {\bibfnamefont {R.}~\bibnamefont {Levine}}, \bibinfo {author} {\bibfnamefont
  {D.}~\bibnamefont {Nagai}}, \ and\ \bibinfo {author} {\bibfnamefont
  {G.}~\bibnamefont {Yepes}},\ }\href@noop {} {\  (\bibinfo {year}
  {2011}{\natexlab{b}})},\ \Eprint {http://arxiv.org/abs/1108.5736}
  {arXiv:1108.5736 [astro-ph.CO]} \BibitemShut {NoStop}%
\bibitem [{\citenamefont {Bouchet}\ \emph {et~al.}(2011)\citenamefont
  {Bouchet}, \citenamefont {Strong}, \citenamefont {Porter}, \citenamefont
  {Moskalenko}, \citenamefont {Jourdain},\ and\ \citenamefont
  {Roques}}]{Bouchet:2011fn}%
  \BibitemOpen
  \bibfield  {author} {\bibinfo {author} {\bibfnamefont {L.}~\bibnamefont
  {Bouchet}}, \bibinfo {author} {\bibfnamefont {A.~W.}\ \bibnamefont {Strong}},
  \bibinfo {author} {\bibfnamefont {T.~A.}\ \bibnamefont {Porter}}, \bibinfo
  {author} {\bibfnamefont {I.~V.}\ \bibnamefont {Moskalenko}}, \bibinfo
  {author} {\bibfnamefont {E.}~\bibnamefont {Jourdain}}, \ and\ \bibinfo
  {author} {\bibfnamefont {J.-P.}\ \bibnamefont {Roques}},\ }\href {\doibase
  10.1088/0004-637X/739/1/29} {\bibfield  {journal} {\bibinfo  {journal}
  {Astrophys. J.}\ }\textbf {\bibinfo {volume} {739}},\ \bibinfo {pages} {29}
  (\bibinfo {year} {2011})},\ \Eprint {http://arxiv.org/abs/1107.0200}
  {arXiv:1107.0200 [astro-ph.HE]} \BibitemShut {NoStop}%
\bibitem [{\citenamefont {Strong}\ \emph {et~al.}(1999)\citenamefont {Strong},
  \citenamefont {Bloemen}, \citenamefont {Diehl}, \citenamefont {Hermsen},\
  and\ \citenamefont {Schoenfelder}}]{Strong:1998ck}%
  \BibitemOpen
  \bibfield  {author} {\bibinfo {author} {\bibfnamefont {A.~W.}\ \bibnamefont
  {Strong}}, \bibinfo {author} {\bibfnamefont {H.}~\bibnamefont {Bloemen}},
  \bibinfo {author} {\bibfnamefont {R.}~\bibnamefont {Diehl}}, \bibinfo
  {author} {\bibfnamefont {W.}~\bibnamefont {Hermsen}}, \ and\ \bibinfo
  {author} {\bibfnamefont {V.}~\bibnamefont {Schoenfelder}},\ }\href@noop {}
  {\bibfield  {journal} {\bibinfo  {journal} {Astrophys. Lett. Commun.}\
  }\textbf {\bibinfo {volume} {39}},\ \bibinfo {pages} {209} (\bibinfo {year}
  {1999})},\ \Eprint {http://arxiv.org/abs/astro-ph/9811211}
  {arXiv:astro-ph/9811211} \BibitemShut {NoStop}%
\bibitem [{\citenamefont {Lehoucq}\ \emph {et~al.}(2009)\citenamefont
  {Lehoucq}, \citenamefont {Casse}, \citenamefont {Casandjian},\ and\
  \citenamefont {Grenier}}]{Lehoucq:2009ge}%
  \BibitemOpen
  \bibfield  {author} {\bibinfo {author} {\bibfnamefont {R.}~\bibnamefont
  {Lehoucq}}, \bibinfo {author} {\bibfnamefont {M.}~\bibnamefont {Casse}},
  \bibinfo {author} {\bibfnamefont {J.~M.}\ \bibnamefont {Casandjian}}, \ and\
  \bibinfo {author} {\bibfnamefont {I.}~\bibnamefont {Grenier}},\ }\href
  {\doibase 10.1051/0004-6361/200911961} {\bibfield  {journal} {\bibinfo
  {journal} {Astron. Astrophys.}\ }\textbf {\bibinfo {volume} {502}},\ \bibinfo
  {pages} {37} (\bibinfo {year} {2009})},\ \Eprint
  {http://arxiv.org/abs/0906.1648} {arXiv:0906.1648 [astro-ph.HE]} \BibitemShut
  {NoStop}%
\bibitem [{\citenamefont {Ackermann}\ \emph {et~al.}(2018)\citenamefont
  {Ackermann} \emph {et~al.}}]{Fermi-LAT:2018pfs}%
  \BibitemOpen
  \bibfield  {author} {\bibinfo {author} {\bibfnamefont {M.}~\bibnamefont
  {Ackermann}} \emph {et~al.} (\bibinfo {collaboration} {Fermi-LAT}),\ }\href
  {\doibase 10.3847/1538-4357/aaac7b} {\bibfield  {journal} {\bibinfo
  {journal} {Astrophys. J.}\ }\textbf {\bibinfo {volume} {857}},\ \bibinfo
  {pages} {49} (\bibinfo {year} {2018})},\ \Eprint
  {http://arxiv.org/abs/1802.00100} {arXiv:1802.00100 [astro-ph.HE]}
  \BibitemShut {NoStop}%
\bibitem [{\citenamefont {Cirelli}\ \emph {et~al.}(2020)\citenamefont
  {Cirelli}, \citenamefont {Fornengo}, \citenamefont {Kavanagh},\ and\
  \citenamefont {Pinetti}}]{Cirelli:2020bpc}%
  \BibitemOpen
  \bibfield  {author} {\bibinfo {author} {\bibfnamefont {M.}~\bibnamefont
  {Cirelli}}, \bibinfo {author} {\bibfnamefont {N.}~\bibnamefont {Fornengo}},
  \bibinfo {author} {\bibfnamefont {B.~J.}\ \bibnamefont {Kavanagh}}, \ and\
  \bibinfo {author} {\bibfnamefont {E.}~\bibnamefont {Pinetti}},\ }\href@noop
  {} {\  (\bibinfo {year} {2020})},\ \Eprint {http://arxiv.org/abs/2007.11493}
  {arXiv:2007.11493 [hep-ph]} \BibitemShut {NoStop}%
\bibitem [{\citenamefont {Hawking}\ \emph {et~al.}(1982)\citenamefont
  {Hawking}, \citenamefont {Moss},\ and\ \citenamefont
  {Stewart}}]{Hawking:1982ga}%
  \BibitemOpen
  \bibfield  {author} {\bibinfo {author} {\bibfnamefont {S.~W.}\ \bibnamefont
  {Hawking}}, \bibinfo {author} {\bibfnamefont {I.~G.}\ \bibnamefont {Moss}}, \
  and\ \bibinfo {author} {\bibfnamefont {J.~M.}\ \bibnamefont {Stewart}},\
  }\href {\doibase 10.1103/PhysRevD.26.2681} {\bibfield  {journal} {\bibinfo
  {journal} {Phys. Rev. D}\ }\textbf {\bibinfo {volume} {26}},\ \bibinfo
  {pages} {2681} (\bibinfo {year} {1982})}\BibitemShut {NoStop}%
\bibitem [{\citenamefont {Garcia-Bellido}\ \emph {et~al.}(1996)\citenamefont
  {Garcia-Bellido}, \citenamefont {Linde},\ and\ \citenamefont
  {Wands}}]{GarciaBellido:1996qt}%
  \BibitemOpen
  \bibfield  {author} {\bibinfo {author} {\bibfnamefont {J.}~\bibnamefont
  {Garcia-Bellido}}, \bibinfo {author} {\bibfnamefont {A.~D.}\ \bibnamefont
  {Linde}}, \ and\ \bibinfo {author} {\bibfnamefont {D.}~\bibnamefont
  {Wands}},\ }\href {\doibase 10.1103/PhysRevD.54.6040} {\bibfield  {journal}
  {\bibinfo  {journal} {Phys. Rev. D}\ }\textbf {\bibinfo {volume} {54}},\
  \bibinfo {pages} {6040} (\bibinfo {year} {1996})},\ \Eprint
  {http://arxiv.org/abs/astro-ph/9605094} {arXiv:astro-ph/9605094} \BibitemShut
  {NoStop}%
\bibitem [{\citenamefont {Kawasaki}\ \emph {et~al.}(2016)\citenamefont
  {Kawasaki}, \citenamefont {Kusenko}, \citenamefont {Tada},\ and\
  \citenamefont {Yanagida}}]{Kawasaki:2016pql}%
  \BibitemOpen
  \bibfield  {author} {\bibinfo {author} {\bibfnamefont {M.}~\bibnamefont
  {Kawasaki}}, \bibinfo {author} {\bibfnamefont {A.}~\bibnamefont {Kusenko}},
  \bibinfo {author} {\bibfnamefont {Y.}~\bibnamefont {Tada}}, \ and\ \bibinfo
  {author} {\bibfnamefont {T.~T.}\ \bibnamefont {Yanagida}},\ }\href {\doibase
  10.1103/PhysRevD.94.083523} {\bibfield  {journal} {\bibinfo  {journal} {Phys.
  Rev. D}\ }\textbf {\bibinfo {volume} {94}},\ \bibinfo {pages} {083523}
  (\bibinfo {year} {2016})},\ \Eprint {http://arxiv.org/abs/1606.07631}
  {arXiv:1606.07631 [astro-ph.CO]} \BibitemShut {NoStop}%
\bibitem [{\citenamefont {Clesse}\ and\ \citenamefont
  {Garc\'\i{}a-Bellido}(2017)}]{Clesse:2016vqa}%
  \BibitemOpen
  \bibfield  {author} {\bibinfo {author} {\bibfnamefont {S.}~\bibnamefont
  {Clesse}}\ and\ \bibinfo {author} {\bibfnamefont {J.}~\bibnamefont
  {Garc\'\i{}a-Bellido}},\ }\href {\doibase 10.1016/j.dark.2016.10.002}
  {\bibfield  {journal} {\bibinfo  {journal} {Phys. Dark Univ.}\ }\textbf
  {\bibinfo {volume} {15}},\ \bibinfo {pages} {142} (\bibinfo {year} {2017})},\
  \Eprint {http://arxiv.org/abs/1603.05234} {arXiv:1603.05234 [astro-ph.CO]}
  \BibitemShut {NoStop}%
\bibitem [{\citenamefont {Kannike}\ \emph {et~al.}(2017)\citenamefont
  {Kannike}, \citenamefont {Marzola}, \citenamefont {Raidal},\ and\
  \citenamefont {Veerm\"ae}}]{Kannike:2017bxn}%
  \BibitemOpen
  \bibfield  {author} {\bibinfo {author} {\bibfnamefont {K.}~\bibnamefont
  {Kannike}}, \bibinfo {author} {\bibfnamefont {L.}~\bibnamefont {Marzola}},
  \bibinfo {author} {\bibfnamefont {M.}~\bibnamefont {Raidal}}, \ and\ \bibinfo
  {author} {\bibfnamefont {H.}~\bibnamefont {Veerm\"ae}},\ }\href {\doibase
  10.1088/1475-7516/2017/09/020} {\bibfield  {journal} {\bibinfo  {journal}
  {JCAP}\ }\textbf {\bibinfo {volume} {09}},\ \bibinfo {pages} {020} (\bibinfo
  {year} {2017})},\ \Eprint {http://arxiv.org/abs/1705.06225} {arXiv:1705.06225
  [astro-ph.CO]} \BibitemShut {NoStop}%
\bibitem [{\citenamefont {Kawasaki}\ \emph {et~al.}(1998)\citenamefont
  {Kawasaki}, \citenamefont {Sugiyama},\ and\ \citenamefont
  {Yanagida}}]{Kawasaki:1997ju}%
  \BibitemOpen
  \bibfield  {author} {\bibinfo {author} {\bibfnamefont {M.}~\bibnamefont
  {Kawasaki}}, \bibinfo {author} {\bibfnamefont {N.}~\bibnamefont {Sugiyama}},
  \ and\ \bibinfo {author} {\bibfnamefont {T.}~\bibnamefont {Yanagida}},\
  }\href {\doibase 10.1103/PhysRevD.57.6050} {\bibfield  {journal} {\bibinfo
  {journal} {Phys. Rev. D}\ }\textbf {\bibinfo {volume} {57}},\ \bibinfo
  {pages} {6050} (\bibinfo {year} {1998})},\ \Eprint
  {http://arxiv.org/abs/hep-ph/9710259} {arXiv:hep-ph/9710259} \BibitemShut
  {NoStop}%
\bibitem [{\citenamefont {Cai}\ \emph {et~al.}(2018)\citenamefont {Cai},
  \citenamefont {Liu},\ and\ \citenamefont {Wang}}]{Cai:2018rqf}%
  \BibitemOpen
  \bibfield  {author} {\bibinfo {author} {\bibfnamefont {R.-G.}\ \bibnamefont
  {Cai}}, \bibinfo {author} {\bibfnamefont {T.-B.}\ \bibnamefont {Liu}}, \ and\
  \bibinfo {author} {\bibfnamefont {S.-J.}\ \bibnamefont {Wang}},\ }\href
  {\doibase 10.1103/PhysRevD.98.043538} {\bibfield  {journal} {\bibinfo
  {journal} {Phys. Rev. D}\ }\textbf {\bibinfo {volume} {98}},\ \bibinfo
  {pages} {043538} (\bibinfo {year} {2018})},\ \Eprint
  {http://arxiv.org/abs/1806.05390} {arXiv:1806.05390 [astro-ph.CO]}
  \BibitemShut {NoStop}%
\bibitem [{\citenamefont {Yoo}\ \emph {et~al.}(2018)\citenamefont {Yoo},
  \citenamefont {Harada}, \citenamefont {Garriga},\ and\ \citenamefont
  {Kohri}}]{Yoo:2018kvb}%
  \BibitemOpen
  \bibfield  {author} {\bibinfo {author} {\bibfnamefont {C.-M.}\ \bibnamefont
  {Yoo}}, \bibinfo {author} {\bibfnamefont {T.}~\bibnamefont {Harada}},
  \bibinfo {author} {\bibfnamefont {J.}~\bibnamefont {Garriga}}, \ and\
  \bibinfo {author} {\bibfnamefont {K.}~\bibnamefont {Kohri}},\ }\href
  {\doibase 10.1093/ptep/pty120} {\bibfield  {journal} {\bibinfo  {journal}
  {PTEP}\ }\textbf {\bibinfo {volume} {2018}},\ \bibinfo {pages} {123E01}
  (\bibinfo {year} {2018})},\ \Eprint {http://arxiv.org/abs/1805.03946}
  {arXiv:1805.03946 [astro-ph.CO]} \BibitemShut {NoStop}%
\bibitem [{\citenamefont {Young}\ and\ \citenamefont
  {Byrnes}(2015)}]{Young:2015kda}%
  \BibitemOpen
  \bibfield  {author} {\bibinfo {author} {\bibfnamefont {S.}~\bibnamefont
  {Young}}\ and\ \bibinfo {author} {\bibfnamefont {C.~T.}\ \bibnamefont
  {Byrnes}},\ }\href {\doibase 10.1088/1475-7516/2015/04/034} {\bibfield
  {journal} {\bibinfo  {journal} {JCAP}\ }\textbf {\bibinfo {volume} {04}},\
  \bibinfo {pages} {034} (\bibinfo {year} {2015})},\ \Eprint
  {http://arxiv.org/abs/1503.01505} {arXiv:1503.01505 [astro-ph.CO]}
  \BibitemShut {NoStop}%
\bibitem [{\citenamefont {Clesse}\ and\ \citenamefont
  {Garc\'\i{}a-Bellido}(2015)}]{Clesse:2015wea}%
  \BibitemOpen
  \bibfield  {author} {\bibinfo {author} {\bibfnamefont {S.}~\bibnamefont
  {Clesse}}\ and\ \bibinfo {author} {\bibfnamefont {J.}~\bibnamefont
  {Garc\'\i{}a-Bellido}},\ }\href {\doibase 10.1103/PhysRevD.92.023524}
  {\bibfield  {journal} {\bibinfo  {journal} {Phys. Rev. D}\ }\textbf {\bibinfo
  {volume} {92}},\ \bibinfo {pages} {023524} (\bibinfo {year} {2015})},\
  \Eprint {http://arxiv.org/abs/1501.07565} {arXiv:1501.07565 [astro-ph.CO]}
  \BibitemShut {NoStop}%
\bibitem [{\citenamefont {Hsu}(1990)}]{Hsu:1990fg}%
  \BibitemOpen
  \bibfield  {author} {\bibinfo {author} {\bibfnamefont {S.~D.~H.}\
  \bibnamefont {Hsu}},\ }\href {\doibase 10.1016/0370-2693(90)90717-K}
  {\bibfield  {journal} {\bibinfo  {journal} {Phys. Lett. B}\ }\textbf
  {\bibinfo {volume} {251}},\ \bibinfo {pages} {343} (\bibinfo {year}
  {1990})}\BibitemShut {NoStop}%
\bibitem [{\citenamefont {La}\ and\ \citenamefont
  {Steinhardt}(1989{\natexlab{a}})}]{La:1989za}%
  \BibitemOpen
  \bibfield  {author} {\bibinfo {author} {\bibfnamefont {D.}~\bibnamefont
  {La}}\ and\ \bibinfo {author} {\bibfnamefont {P.~J.}\ \bibnamefont
  {Steinhardt}},\ }\href {\doibase 10.1103/PhysRevLett.62.376} {\bibfield
  {journal} {\bibinfo  {journal} {Phys. Rev. Lett.}\ }\textbf {\bibinfo
  {volume} {62}},\ \bibinfo {pages} {376} (\bibinfo {year}
  {1989}{\natexlab{a}})},\ \bibinfo {note} {[Erratum: Phys.Rev.Lett. 62, 1066
  (1989)]}\BibitemShut {NoStop}%
\bibitem [{\citenamefont {La}\ and\ \citenamefont
  {Steinhardt}(1989{\natexlab{b}})}]{La:1989st}%
  \BibitemOpen
  \bibfield  {author} {\bibinfo {author} {\bibfnamefont {D.}~\bibnamefont
  {La}}\ and\ \bibinfo {author} {\bibfnamefont {P.~J.}\ \bibnamefont
  {Steinhardt}},\ }\href {\doibase 10.1016/0370-2693(89)90890-3} {\bibfield
  {journal} {\bibinfo  {journal} {Phys. Lett. B}\ }\textbf {\bibinfo {volume}
  {220}},\ \bibinfo {pages} {375} (\bibinfo {year}
  {1989}{\natexlab{b}})}\BibitemShut {NoStop}%
\bibitem [{\citenamefont {La}\ \emph {et~al.}(1989)\citenamefont {La},
  \citenamefont {Steinhardt},\ and\ \citenamefont {Bertschinger}}]{La:1989pn}%
  \BibitemOpen
  \bibfield  {author} {\bibinfo {author} {\bibfnamefont {D.}~\bibnamefont
  {La}}, \bibinfo {author} {\bibfnamefont {P.~J.}\ \bibnamefont {Steinhardt}},
  \ and\ \bibinfo {author} {\bibfnamefont {E.~W.}\ \bibnamefont
  {Bertschinger}},\ }\href {\doibase 10.1016/0370-2693(89)90205-0} {\bibfield
  {journal} {\bibinfo  {journal} {Phys. Lett. B}\ }\textbf {\bibinfo {volume}
  {231}},\ \bibinfo {pages} {231} (\bibinfo {year} {1989})}\BibitemShut
  {NoStop}%
\bibitem [{\citenamefont {Weinberg}(1989)}]{PhysRevD.40.3950}%
  \BibitemOpen
  \bibfield  {author} {\bibinfo {author} {\bibfnamefont {E.~J.}\ \bibnamefont
  {Weinberg}},\ }\href {\doibase 10.1103/PhysRevD.40.3950} {\bibfield
  {journal} {\bibinfo  {journal} {Phys. Rev. D}\ }\textbf {\bibinfo {volume}
  {40}},\ \bibinfo {pages} {3950} (\bibinfo {year} {1989})}\BibitemShut
  {NoStop}%
\bibitem [{\citenamefont {Steinhardt}\ and\ \citenamefont
  {Accetta}(1990)}]{Steinhardt:1990zx}%
  \BibitemOpen
  \bibfield  {author} {\bibinfo {author} {\bibfnamefont {P.~J.}\ \bibnamefont
  {Steinhardt}}\ and\ \bibinfo {author} {\bibfnamefont {F.~S.}\ \bibnamefont
  {Accetta}},\ }\href {\doibase 10.1103/PhysRevLett.64.2740} {\bibfield
  {journal} {\bibinfo  {journal} {Phys. Rev. Lett.}\ }\textbf {\bibinfo
  {volume} {64}},\ \bibinfo {pages} {2740} (\bibinfo {year}
  {1990})}\BibitemShut {NoStop}%
\bibitem [{\citenamefont {Holman}\ \emph {et~al.}(1990)\citenamefont {Holman},
  \citenamefont {Kolb},\ and\ \citenamefont {Wang}}]{Holman:1990wq}%
  \BibitemOpen
  \bibfield  {author} {\bibinfo {author} {\bibfnamefont {R.}~\bibnamefont
  {Holman}}, \bibinfo {author} {\bibfnamefont {E.~W.}\ \bibnamefont {Kolb}}, \
  and\ \bibinfo {author} {\bibfnamefont {Y.}~\bibnamefont {Wang}},\ }\href
  {\doibase 10.1103/PhysRevLett.65.17} {\bibfield  {journal} {\bibinfo
  {journal} {Phys. Rev. Lett.}\ }\textbf {\bibinfo {volume} {65}},\ \bibinfo
  {pages} {17} (\bibinfo {year} {1990})}\BibitemShut {NoStop}%
\bibitem [{\citenamefont {Khlopov}\ and\ \citenamefont
  {Polnarev}(1980)}]{Khlopov:1980mg}%
  \BibitemOpen
  \bibfield  {author} {\bibinfo {author} {\bibfnamefont {M.~Y.}\ \bibnamefont
  {Khlopov}}\ and\ \bibinfo {author} {\bibfnamefont {A.~G.}\ \bibnamefont
  {Polnarev}},\ }\href {\doibase 10.1016/0370-2693(80)90624-3} {\bibfield
  {journal} {\bibinfo  {journal} {Phys. Lett. B}\ }\textbf {\bibinfo {volume}
  {97}},\ \bibinfo {pages} {383} (\bibinfo {year} {1980})}\BibitemShut
  {NoStop}%
\bibitem [{\citenamefont {Caputo}\ \emph {et~al.}(2019)\citenamefont {Caputo}
  \emph {et~al.}}]{McEnery:2019tcm}%
  \BibitemOpen
  \bibfield  {author} {\bibinfo {author} {\bibfnamefont {R.}~\bibnamefont
  {Caputo}} \emph {et~al.} (\bibinfo {collaboration} {AMEGO}),\ }\href@noop {}
  {\  (\bibinfo {year} {2019})},\ \Eprint {http://arxiv.org/abs/1907.07558}
  {arXiv:1907.07558 [astro-ph.IM]} \BibitemShut {NoStop}%
\bibitem [{\citenamefont {De~Angelis}\ \emph {et~al.}(2017)\citenamefont
  {De~Angelis} \emph {et~al.}}]{DeAngelis:2016slk}%
  \BibitemOpen
  \bibfield  {author} {\bibinfo {author} {\bibfnamefont {A.}~\bibnamefont
  {De~Angelis}} \emph {et~al.} (\bibinfo {collaboration} {e-ASTROGAM}),\ }\href
  {\doibase 10.1007/s10686-017-9533-6} {\bibfield  {journal} {\bibinfo
  {journal} {Exper. Astron.}\ }\textbf {\bibinfo {volume} {44}},\ \bibinfo
  {pages} {25} (\bibinfo {year} {2017})},\ \Eprint
  {http://arxiv.org/abs/1611.02232} {arXiv:1611.02232 [astro-ph.HE]}
  \BibitemShut {NoStop}%
\bibitem [{\citenamefont {Sobrinho}\ and\ \citenamefont
  {Augusto}(2014)}]{Sobrinho:2014vka}%
  \BibitemOpen
  \bibfield  {author} {\bibinfo {author} {\bibfnamefont {J.~L.~G.}\
  \bibnamefont {Sobrinho}}\ and\ \bibinfo {author} {\bibfnamefont
  {P.}~\bibnamefont {Augusto}},\ }\href {\doibase 10.1093/mnras/stu786}
  {\bibfield  {journal} {\bibinfo  {journal} {Mon. Not. Roy. Astron. Soc.}\
  }\textbf {\bibinfo {volume} {441}},\ \bibinfo {pages} {2878} (\bibinfo {year}
  {2014})},\ \Eprint {http://arxiv.org/abs/1406.1785} {arXiv:1406.1785
  [astro-ph.CO]} \BibitemShut {NoStop}%
\bibitem [{\citenamefont {Ray}\ \emph {et~al.}(2021)\citenamefont {Ray},
  \citenamefont {Laha}, \citenamefont {Mu\~noz},\ and\ \citenamefont
  {Caputo}}]{Ray:2021mxu}%
  \BibitemOpen
  \bibfield  {author} {\bibinfo {author} {\bibfnamefont {A.}~\bibnamefont
  {Ray}}, \bibinfo {author} {\bibfnamefont {R.}~\bibnamefont {Laha}}, \bibinfo
  {author} {\bibfnamefont {J.~B.}\ \bibnamefont {Mu\~noz}}, \ and\ \bibinfo
  {author} {\bibfnamefont {R.}~\bibnamefont {Caputo}},\ }\href@noop {} {\
  (\bibinfo {year} {2021})},\ \Eprint {http://arxiv.org/abs/2102.06714}
  {arXiv:2102.06714 [astro-ph.CO]} \BibitemShut {NoStop}%
\bibitem [{\citenamefont {Carr}\ \emph
  {et~al.}(2020{\natexlab{b}})\citenamefont {Carr}, \citenamefont {Kuhnel},\
  and\ \citenamefont {Visinelli}}]{Carr:2020mqm}%
  \BibitemOpen
  \bibfield  {author} {\bibinfo {author} {\bibfnamefont {B.}~\bibnamefont
  {Carr}}, \bibinfo {author} {\bibfnamefont {F.}~\bibnamefont {Kuhnel}}, \ and\
  \bibinfo {author} {\bibfnamefont {L.}~\bibnamefont {Visinelli}},\ }\href@noop
  {} {\  (\bibinfo {year} {2020}{\natexlab{b}})},\ \Eprint
  {http://arxiv.org/abs/2011.01930} {arXiv:2011.01930 [astro-ph.CO]}
  \BibitemShut {NoStop}%
\bibitem [{\citenamefont {Adamek}\ \emph {et~al.}(2019)\citenamefont {Adamek},
  \citenamefont {Byrnes}, \citenamefont {Gosenca},\ and\ \citenamefont
  {Hotchkiss}}]{Adamek:2019gns}%
  \BibitemOpen
  \bibfield  {author} {\bibinfo {author} {\bibfnamefont {J.}~\bibnamefont
  {Adamek}}, \bibinfo {author} {\bibfnamefont {C.~T.}\ \bibnamefont {Byrnes}},
  \bibinfo {author} {\bibfnamefont {M.}~\bibnamefont {Gosenca}}, \ and\
  \bibinfo {author} {\bibfnamefont {S.}~\bibnamefont {Hotchkiss}},\ }\href
  {\doibase 10.1103/PhysRevD.100.023506} {\bibfield  {journal} {\bibinfo
  {journal} {Phys. Rev. D}\ }\textbf {\bibinfo {volume} {100}},\ \bibinfo
  {pages} {023506} (\bibinfo {year} {2019})},\ \Eprint
  {http://arxiv.org/abs/1901.08528} {arXiv:1901.08528 [astro-ph.CO]}
  \BibitemShut {NoStop}%
\bibitem [{\citenamefont {Beacom}\ \emph {et~al.}(2005)\citenamefont {Beacom},
  \citenamefont {Bell},\ and\ \citenamefont {Bertone}}]{Beacom:2004pe}%
  \BibitemOpen
  \bibfield  {author} {\bibinfo {author} {\bibfnamefont {J.~F.}\ \bibnamefont
  {Beacom}}, \bibinfo {author} {\bibfnamefont {N.~F.}\ \bibnamefont {Bell}}, \
  and\ \bibinfo {author} {\bibfnamefont {G.}~\bibnamefont {Bertone}},\ }\href
  {\doibase 10.1103/PhysRevLett.94.171301} {\bibfield  {journal} {\bibinfo
  {journal} {Phys. Rev. Lett.}\ }\textbf {\bibinfo {volume} {94}},\ \bibinfo
  {pages} {171301} (\bibinfo {year} {2005})},\ \Eprint
  {http://arxiv.org/abs/astro-ph/0409403} {arXiv:astro-ph/0409403} \BibitemShut
  {NoStop}%
\end{thebibliography}%

\end{document}